\documentclass{emulateapj}

\usepackage{graphicx, epsf}
\usepackage{amsmath, amssymb}

\newcommand{\velunits}{~$\rm km~s^{-1}$}

\newcommand{\vph}{\ensuremath{v_{\mathrm{ph}}}}

\def\vec#1{\ensuremath{\mathbf{#1}}}

\shorttitle{Period ratio of magnetic slabs}
\shortauthors{Li et al.}

\begin{document}
\title{The period ratio for standing kink and sausage modes in { solar} structures with siphon flow. I. {magnetized} slabs}
\author{Bo Li}
\affil{Shandong Provincial Key Laboratory of Optical Astronomy \& Solar-Terrestrial Environment, 
  School of Space Science and Physics, Shandong University at Weihai, 264209, Weihai, PR China}
\email{bbl@sdu.edu.cn}

\author{Shadia Rifai Habbal}
\affil{Institute for Astronomy, University of Hawaii, HI 96822, USA}

\and
\author{Yanjun Chen}
\affil{School of Space Science and Physics, Shandong University at Weihai, 264209, Weihai, PR China}

\begin{abstract}
{
In the applications of solar magneto-seismology(SMS), employing the ratio of the period of the fundamental mode to twice the one of its first overtone, $P_1/2P_2$, plays an important role. We examine how field-aligned flows affect the dispersion properties, and hence the period ratios, of standing modes supported by magnetic slabs in the solar atmosphere. We numerically solve the dispersion relations and devise a graphic means to construct standing modes. For coronal slabs, we find that the flow effects are significant, for the fast kink and sausage modes alike. For the kink ones, they may reduce $P_1/2P_2$ by up to 23\% compared with the static case, and the minimum allowed $P_1/2P_2$ can fall below the lower limit analytically derived for static slabs. For the sausage modes, while introducing the flow reduces $P_1/2P_2$ by typically $\lesssim 5$\% relative to the static case, it significantly increases the threshold aspect ratio only above which standing sausage modes can be supported, meaning that their 
detectability is restricted to even wider slabs. In the case of photospheric slabs, the flow effect is not as strong. However, standing modes are distinct from the coronal case in that standing kink modes show a $P_1/2P_2$ that deviates from unity even for a zero-width slab, while standing sausage modes no longer suffer from a threshold aspect ratio. We conclude that transverse structuring in plasma density and flow speed should be considered in seismological applications of multiple periodicities to solar atmospheric structures.
}
\end{abstract}

\keywords{magnetohydrodynamics (MHD) -- Sun: corona -- Sun: oscillations -- waves}

\section{INTRODUCTION}

A rich variety of magnetohydrodynamic (MHD) waves and oscillations
   have been detected
   in the structured solar { atmosphere}
   \citep[e.g., the reviews by][]{2004psci.book.....A,2005LRSP....2....3N,2009SSRv..149....1N,2011SSRv..158..167E,2012RSPTA.370.3193D}.
While identifying waves/oscillations was possible with radio and optical
   instruments prior to the 1990s~\citep[][and references therein]{1984ApJ...279..857R},
   the majority of such observations emerged only after the advent of the Solar
   and Heliospheric Observatory (SOHO),
   the Transition Region and Corona Explorer (TRACE),
   and later Hinode,
   in conjunction with the availability
   of ground-based instruments with very high temporal and spatial resolution
   such as the Dutch Open Telescope (DOT)
   and the Swedish Solar Telescope (SST).
For instance, imaging and spectroscopic instruments on SOHO, along with the H$_\alpha$
   measurements with SST and DOT have shown that MHD oscillations abound in
   prominences - cool and dense clouds suspended in the corona
   \citep[][and references therein]{2012LRSP....9....2A}.
Instruments with high spatial resolution like the Solar Optical Telescope (SOT)
   on board Hinode have detected transverse oscillatory loop-like
   structures accompanied by downflows termed ``coronal rains''
   \citep{2011ApJ...736..121A}.
Hinode/SOT has also detected transverse oscillations in spicules - chromospheric materials
   isolated from a coronal environment
   \citep{2007Sci...318.1574D,2009A&A...497..525H}.
Using the Ultraviolet Coronagraph Spectrometer (UVCS) data,
   \citet{2004ApJ...605..521M} detected significant periodic oscillations in H I Ly$\alpha$ intensities in
   a coronal hole, the quiet Sun and a streamer.
{Coronal loops offer by far the most samples of MHD oscillations, among which 
   the majority seems to be standing transverse oscillations
   triggered by flares~\citep[e.g.,][]{1999ApJ...520..880A, 1999Sci...285..862N}.
}

Identifying the observed oscillations with a particular MHD mode collectively
   supported by a magnetized structure is physically crucial but proves difficult.
This is best illustrated by the substantial controversy associated with interpreting
   the observed transverse oscillations in terms of Alfv\'en
   waves~\citep[for the most recent review, see][]{2012SSRv..tmp...94M}.
This happens with both coronal~\citep{2007Sci...317.1192T}
   and spicule measurements~\citep{2007Sci...318.1574D},
   the interpretation of which
   as Alfv\'en waves
   was suggested to be inconsistent with theory by~\citet{2007Sci...318.1572E}
   and later by~\citet{2008ApJ...676L..73V}.
The first unambiguous detection of Alfv\'en waves in the solar atmosphere
   is due to~\citet{2009Sci...323.1582J}
   where H$\alpha$ observations with SST of a group of magnetic bright points are analyzed.
Less controversial is the identification of fast and slow waves.
Consider first the coronal case.
Flare-triggered transverse oscillations supported by loops are usually  
   fast kink modes~\citep[e.g.,][]{1999ApJ...520..880A, 1999Sci...285..862N, 2008A&A...487L..17V, 2009ApJ...698..397V}.
On the other hand, fast sausage modes were also inferred
   albeit much less abundant~\citep{2003A&A...412L...7N, 2005A&A...439..727M}.
As for the slow waves, they are frequently observed and 
   appear both in the form of standing \citep[][and references therein]{2011SSRv..158..397W}
   and as propagating ones \citep[see ][for comprehensive reviews]{2006RSPTA.364..461D, 2007SoPh..246....3B,2009SSRv..149...65D}.  
When it comes to atmospheric parts other than the corona,
   sausage waves are directly observed in the photosphere~\citep{2011ApJ...729L..18M}
   and chromosphere~\citep{2012NatCom..3.1315M}.
Note that in the fine structures examined in the latter study, 
   these fast sausage waves, manifested as intensity oscillations,
   appear simultaneously with fast kink waves, manifested in the form of transverse displacements.  
That periodic longitudinal and transverse oscillatory motions may accompany each other
   was first shown by~\citet{2008A&A...489L..49E} for Hinode loops.
At this point, it should be noted that adopting a helioseismic approach by working with the power spectra of, say,
   the Doppler shift time series,  
   can greatly improve the capability for wave identification,
   as was advocated by~\citet{2007A&A...462..331T}.

Once identified, the measured properties of waves and oscillations can then be combined with MHD theory
   to yield parameters of the solar atmosphere that are otherwise difficult to measure.   
{The idea behind this was put forward as early as in the 1970's following the pioneering work
   by~\citet{1970PASJ...22..341U} and \citet{ZS75}.
Along with others, Roberts and co-workers~\citep[e.g.,][]{1982SoPh...76..239E,1983SoPh...88..179E,1984ApJ...279..857R} expressed the idea
   in the mathematical form now in routine use.
This powerful diagnostic capability, 
}
   initially termed ``coronal seismology'',
   may be properly called ``solar magneto-seismology'' (SMS), given that
   both observations and seismological applications extend well beyond the solar corona
   (for recent reviews, see~\citeauthor{2000SoPh..193..139R}~\citeyear{2000SoPh..193..139R},
   \citeyear{2008IAUS..247....3R};
   \citeauthor{2005LRSP....2....3N}~\citeyear{2005LRSP....2....3N};
   \citeauthor{2009SSRv..149..199R}~\citeyear{2009SSRv..149..199R}).
The parameter that is perhaps most frequently inferred is the magnetic field strength.
For instance, measuring the periodic flare-generated transverse displacements in several TRACE loops
   leads to concrete estimates of the coronal magnetic
   field strength~\citep{2001A&A...372L..53N}.
This practice was also applied to loops observed by
   Hinode/SOT~\citep[e.g.,][]{2008A&A...489L..49E,2008A&A...482L...9O,2008A&A...487L..17V},
   and recently by the Atmospheric Imaging Assembly (AIA) on board
   the Solar Dynamic Observatory (SDO)~\citep{2012A&A...537A..49W}.
Certainly SMS can offer more. 
As a matter of fact, slow waves running in Active Region loops measured by
   the Solar Terrestrial Relations Observatory (STEREO)
   enabled~\citet{2009ApJ...697.1674M} to infer loop temperatures
   via the measured propagation speeds.
Slow waves measured by Hinode also helped derive the effective adiabatic
   index~\citep{2011ApJ...727L..32V}.

{Multiple periodicities detected in some oscillating structures
   are playing an increasingly important role
   in solar magneto-seismology~\citep[see][for recent reviews]{2009SSRv..149....3A, 2009SSRv..149..199R}.}
They may have disparate timescales, differing by nearly a factor of $10$
   as shown in the light curve data obtained by the LYRA irradiance experiment,
   indicative of the coexistence of slow
   and fast modes~\citep{2011ApJ...740...90V}.
This coexistence can help derive the plasma beta and density contrast of
   the flare loops where the oscillations are found.
They may have comparable timescales, favoring an interpretation of the coexistence
   of a fundamental mode with its overtones.
Two~\citep[e.g.,][]{2004SoPh..223...77V,2007A&A...473..959V}
   and three periodicities~\citep{2007ApJ...664.1210D, 2009A&A...508.1485V,2009A&A...493..259I}
   have been found.
One thing peculiar with the multiple periodicities is that
   the ratio between the period of the fundamental and twice the period of its first overtone, $P_1/2P_2$,
   deviates in general from unity.
For instance, the loops labeled C and D in the TRACE 171\AA\ images on 15 April 2001
   exhibit a $P_1/2P_2$ of 0.91 and 0.82, respectively~\citep{2004SoPh..223...77V}.
A further study using similar TRACE 171\AA\ loops on 23 Nov 1998
    yields a $P_1/2P_2$ of 0.9~\citep{2007A&A...473..959V}.
This is not expected for kink modes supported by a longitudinally uniform loop with tiny aspect ratios.
Hence the deviation of $P_1/2P_2$ from $1$
    may signify some longitudinal structuring.
Actually~\citet{2005ApJ...624L..57A} come up with a one-to-one look-up diagram
    where the density scale height along the loop
    in units of the loop length
    can be readily deduced once $1-P_1/2P_2$ is known.
Fundamental or global sausage modes with their first overtones were also found.
For example, flare-associated quasi-periodic pulsations measured with
   the Nobeyama Radioheliograph (NoRH) on 12 Jan 2000 contain a global sausage mode
   with period $P_1 \sim 14-17$~s, and a higher
    harmonic with $P_2 \sim 8-11$~s~\citep{2003A&A...412L...7N, 2005A&A...439..727M}.
Moreover, high spatial resolution H$_\alpha$ images of oscillating cool post-flare loops measured with
   the Solar Tower Telescope at Aryabhatta Research Institute of Observational Sciences (ARIES)
   yield a $P_1 \approx 587$~s and a $P_2 \approx 349$~s~\citep{2008MNRAS.388.1899S}.

Flows are ubiquitous in {the solar atmosphere}~\citep[e.g.,][]{2004psci.book.....A}
   and have been found in oscillating structures~\citep[e.g.,][]{2008A&A...482L...9O, 2008MNRAS.388.1899S}.
{Take the corona for instance.}
The frequently found flow speeds, usually $\lesssim 100$~\velunits\ and hence
   well below the Alfv\'{e}n speed, have only weak effects on the periods of the fundamental
   kink mode and its overtone supported by thin loops~\citep[e.g.,][]{2010SoPh..267..377R}.
Therefore it is suggested that the effect of flow may be neglected
   in the seismological applications of $1-P_1/2P_2$ {to the corona}.
However, the flow speeds may not be always small.
Although less frequent, flow speeds reaching the Alfv\'{e}nic range ($\sim 10^3$\velunits)
   have been seen associated with explosive events~\citep[e.g.,][]{2003SoPh..217..267I,2005A&A...438.1099H}.
Taking into account a siphon flow of the order of the Alfv\'{e}n speed may lead to
   significant revisions to the loop parameters deduced from seismology.
For the standing kink mode measured with TRACE and SoHO on 15 Sept 2001~\citep{2010ApJ...717..458V},
   \citet{2011ApJ...729L..22T} show that
   neglecting the flows leads to an underestimation
   of the loop magnetic field by up to a factor of three!
Such a revision is important in its own right, and also important in the context of coronal heating
   because the energy flux density carried by the transverse waves is proportional to
   the magnetic field strength.

Given the importance of multiple periodicities in the context of {solar magneto-seismology},
   and the extent to which a flow may affect the seismological applications,
   it seems natural to examine further how the flows affect
   multiple periodicities from the theoretical perspective.
In the case of cylindrical geometry, this has been done by~\citet{2010SoPh..267..377R} in the
   thin-tube approximation.
Here we consider a slab geometry first, and leave the study of coronal cylinders with
   finite aspect ratio to a future publication.
By doing so, we will complement and expand the only study concerning the period ratios of a {coronal}
   slab~\citep{2011A&A...526A..75M}, which shows that the transverse structuring may contribute
   to the deviation of the period ratio from $1$.
This is true even for relatively thin slabs, and is more pronounced for high density contrasts.
Extending the static slab study by~\citet{2011A&A...526A..75M} to incorporate flows,
   we will examine how the flows affect the dispersion properties and hence the period ratios
   of both standing kink and sausage modes.

This paper is organized as follows.
In section~\ref{sec_DR_Oview}, we briefly describe the slab dispersion relation.
Section~\ref{sec_Cor_Pratio} {deals with coronal slabs by first
   giving an overview of their dispersion diagrams,
   then describing a graphical means to derive
   the period ratios,
   and also examining} in detail how the flow affects the period ratios for
   the standing kink and sausage modes.
{The same approach is also applied to isolated photospheric slabs,
   the results thus derived are given in section~\ref{sec_Pho_Pratio}.} 
Finally, section~\ref{sec_summary} summarizes the results, ending with some
   concluding remarks.

\section{Slab Dispersion Relation}
\label{sec_DR_Oview}

\begin{figure}
\centering
\epsscale{1.}
\plotone{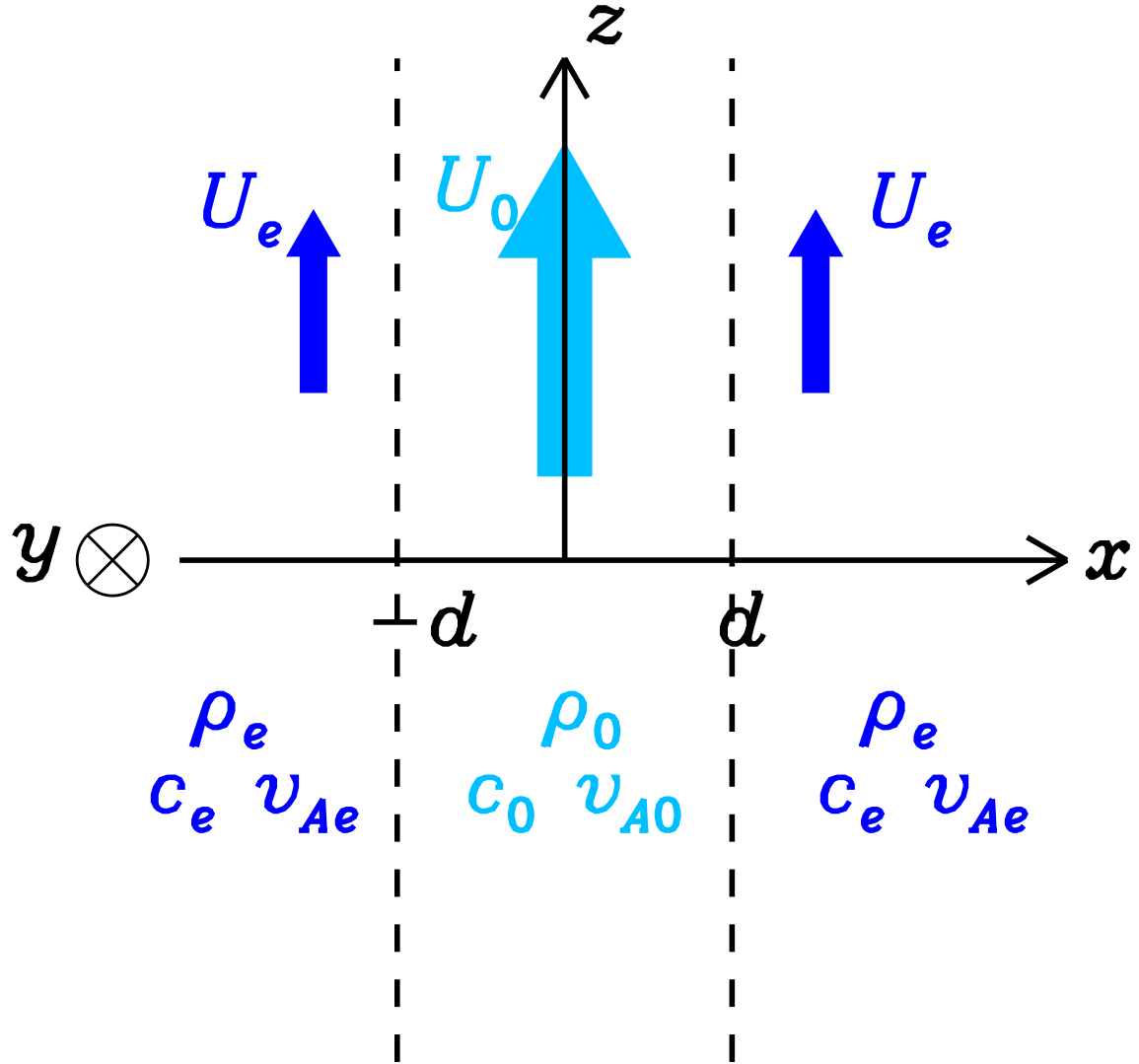}
 \caption{Schematic diagram illustrating the magnetic slab (denoted by subscript $0$)
    and its environment (subscript $e$).
 The variables $\rho_i, c_i, v_{Ai}$ and $U_i$ ($i=0, e$) represent the mass density,
    adiabatic sound speed, Alfv\'en speed, and the flow speed, respectively.
}
 \label{fig_slab_illus}
\end{figure}

Consider a slab of half-width $d$ with time-independent siphon flows.
As illustrated in Fig.\ref{fig_slab_illus},
    the slab is infinite in the $y-$ and $z-$ directions,
    but is bordered by two interfaces $x=\pm d$.
Let the subscripts $0$ and $e$ denote the otherwise uniform parameters
     inside and external to the slab, respectively.
The background magnetic fields ($\vec{B}_0$ and $\vec{B}_e$), together with the
    flow velocities ($\vec{U}_0$
    and $\vec{U}_e$), are in the $z-$direction.
Let $\rho$ and $p$ denote the mass density and thermal pressure.
From the force balance condition across the interfaces it follows that
\begin{eqnarray}
\frac{\rho_e}{\rho_0} = \frac{2 c_0^2 + \gamma v_{A0}^2}{2 c_e^2 + \gamma v_{Ae}^2} ,
\label{eq_rhoe0}
\end{eqnarray}
    where $\gamma=5/3$ is the adiabatic index,
    $c=\sqrt{\gamma p/\rho}$ is the adiabatic sound
    and $v_A=\sqrt{B^2/4\pi\rho}$ the Alfv\'{e}n speed.
To proceed, it proves convenient to introduce the tube speed
\begin{eqnarray}
 c_{Ti}^2 = \frac{c_{i}^2 v_{A i}^2}{c_{i}^2 + v_{A i}^2} ,
 \label{eq_def_ct}
\end{eqnarray}
   where $i=0, e$.

The dispersion relation (DR) for the waves supported by a slab with flow has been examined
    by a number of authors~\citep[e.g.,][]{1995SoPh..159..213N, 2000A&A...359.1211J, 2003PhPl...10.4463M}
    (also see~\citeauthor{2003SoPh..217..199T}~\citeyear{2003SoPh..217..199T} for the
    waves supported by cylinders with flows).
Restricting oneself to propagation in the $x-z$ plane, and to linear perturbations, one may
    adopt the ansatz that any perturbation $\delta f(x, z; t)$ to the equilibrium $f(x)$ is in the form
\begin{eqnarray}
 \delta f(x, z; t) = \mathrm{Re}\left\{\tilde{f}(x)\exp\left[-i\left(\omega t -
kz\right)\right]\right\} ,
\label{eq_wav_ansatz}
\end{eqnarray}
  where $\mathrm{Re}(\cdots)$ means taking the real part of the function.
The phase speed $\vph$ is defined as $\vph=\omega/k$.
Of critical importance is the variable
\begin{eqnarray}
 m_i^2 = k^2 \frac{\left[c_i^2 - \left(\vph -U_i\right)^2\right]\left[v_{Ai}^2 - \left(\vph -U_i\right)^2\right]}
   {\left(c_i^2 + v_{A i}^2\right) \left[c_{T i}^2 - \left(\vph -U_i \right)^2\right]} ,
\label{eq_def_m2}
\end{eqnarray}
where $i=0, e$.
To ensure the waves external to the slab are evanescent, one has to require a positive $m_e^2$, which translates into
\begin{eqnarray}
 c_{\mathrm{m},e}^2 < (\vph-U_e)^2 < c_{\mathrm{M},e}^2 \text{ or }
    (\vph-U_e)^2 < c_{Te}^2 ,
\end{eqnarray}
where $c_{\mathrm{m},e} = \mathrm{min}(c_e, v_{Ae})$
 and $c_{\mathrm{M},e} = \mathrm{max}(c_e, v_{Ae})$.
On the other hand, the signs of $m_0^2$ determine the spatial profiles of the perturbations
    in the transverse (i.e., $x-$) direction.
For $m_0^2 <0$ ($m_0^2 >0$), the modes are  body (surface) waves,
    corresponding to an oscillatory (a spatially decaying) $x-$ dependence
    inside the slab.
Evidently, body waves correspond to
\begin{eqnarray}
(\vph-U_0)^2 > c_{\mathrm{M}, 0}^2  \text{ or }
    c_{T0}^2<(\vph-U_0)^2 < c_{\mathrm{m}, 0}^2 ,
\end{eqnarray}
whereas surface waves take up the rest of the possible $\vph$.

Plugging Eq.(\ref{eq_wav_ansatz}) into the linearized, ideal MHD equations, one may derive
the DR by requiring {the following boundary conditions be satisfied:  
    First, the transverse (i.e., $x-$) component of the Lagrangian displacement $\xi_x$ is continuous at $x=\pm d$.
    Second, so is the total pressure $\delta p_T$.}
{The DR reads~\citep{1995SoPh..159..213N}
 {(see also~\citeauthor{1982SoPh...76..239E}~\citeyear{1982SoPh...76..239E} for its static counterpart)}  
\begin{eqnarray}
 \frac{\rho_e}{\rho_0}\frac{m_0}{|m_e|}
   \frac{\left[v_{Ae}^2 - \left(\vph - U_e\right)^2\right]}{\left[v_{A0}^2 - \left(\vph - U_0\right)^2\right]}
 =-\left\{\begin{array}{c}
           \tanh \\
           \coth
          \end{array}
\right\}\left(m_0 d\right)
\label{eq_DR_surface}
\end{eqnarray}
for surface waves,
and
\begin{eqnarray}
 \frac{\rho_e}{\rho_0}\frac{n_0}{|m_e|}
   \frac{\left[v_{Ae}^2 - \left(\vph - U_e\right)^2\right]}{\left[v_{A0}^2 - \left(\vph - U_0\right)^2\right]}
 =\left\{\begin{array}{c}
           -\tan \\
           \cot
          \end{array}
\right\}\left(n_0 d\right)
\label{eq_DR_body}
\end{eqnarray}
for body waves ($n_0^2 = -m_0^2 >0$).
Furthermore, the upper (lower) case is for kink (sausage) modes.
Note that the absolute value of $m_e$ appears to ensure the waves are trapped, i.e., evanescent outside the slab.
}

{It proves necessary to examine also
    the relative importance of transverse displacement
    versus the density fluctuation.
This is evaluated by $X \equiv \left.(\tilde{\rho}/\rho_0 )/(\tilde{\xi}_x/d)\right|_{x=d}$, which reads
\begin{eqnarray}
X =\left. \frac{d m_0^2 (\omega - k U_0)^2}{\left[k^2 c_0^2 - (\omega - k U_0)^2\right]} \frac{\tilde{p}_T}{d\tilde{p}_T/dx} \right |_{x=d} .
\label{eq_ratio_rho_xi}
 \end{eqnarray}
For body waves, $(d\tilde{p}_T/dx)/\tilde{p}_T$ inside the slab equals $n_0 \cot(n_0 x)$ for a kink wave, and
   $-n_0 \tan(n_0 x)$ for a sausage one, leading to
\begin{eqnarray}
X=
 \frac{(\vph - U_0)^2}{(\vph - U_0)^2 - c_0^2}
 \left\{\begin{array}{l}
  (n_0 d)\tan(n_0 d) \\
		\hskip 0.5cm \text{kink},  \\
- (n_0 d)\cot(n_0 d) \\
		\hskip 0.5cm  \text{sausage}.
		\end{array} \right.
\label{eq_Bdy_rhoOxi}
\end{eqnarray}
For surface waves,  $(d\tilde{p}_T/dx)/\tilde{p}_T$ inside the slab is  $m_0 \coth(m_0 x)$ for a kink wave,
    and $m_0 \tanh(m_0 x)$ for a sausage one, leading to
\begin{eqnarray}
X =
  \frac{(\vph - U_0)^2}{c_0^2 - (\vph - U_0)^2}
  \left\{\begin{array}{l}
  (m_0 d)\tanh(m_0 d) \\
		\hskip 0.5cm \text{kink},  \\
  (m_0 d)\coth(m_0 d) \\
		\hskip 0.5cm  \text{sausage}.
		\end{array} \right.
\end{eqnarray}
}

{Before presenting the solutions to the dispersion relations~(\ref{eq_DR_surface}) and (\ref{eq_DR_body}), 
   let us demonstrate the following symmetry properties.
\begin{enumerate}
 \item If $[\vph, k; U_0, U_e]$ represents a solution to the DR, then 
   so does $[\vph, -k; U_0, U_e]$.
This is true because both $m_0^2$ and $m_e^2$ depend
   on $k$ only via $k^2$, and $m_e$ appears in the absolute value bars.
Therefore changing the sign of $k$ changes the sign of $m_0$, and consequently 
   simultaneously changes the signs of both sides of the DRs. 
The end result is that, the dispersion diagram expressing
   $\vph$ as a function of $k$ is symmetric about the $\vph-$ axis
   and one needs only to consider the 1st and 4th quadrants, as will be done
   in Figs.\ref{fig_Cor_DR} and \ref{fig_Pho_DR}.
For more details, please see appendix.
\item When the external medium is at rest, 
   if $[\vph, k; U_0]$ is a solution, then
   so is $[-\vph, k; -U_0]$ (see Eq.(\ref{eq_def_m2}) with $U_e = 0$).
It follows that
   the dispersion diagram for some positive $U_0$ is a reflection about the $k-$ axis
   of the diagram for $-U_0$.
Actually this tendency was clear already in Figs.4 and 5 in \citet{2003SoPh..217..199T}, 
   even though the negative flow speed adopted in constructing Fig.4 therein
    is slightly smaller in magnitude than the positive one used in generating Fig.5.
More discussions on this are given in the appendix.
\end{enumerate}
We stress that if $U_e$ is not zero, then symmetry 2 does not hold.
The dispersion behavior, or rather the wave behavior in general, 
   critically depends on the signs of $U_e$
   and $U_0$~\citep[see the recent review by][]{2011SSRv..158..505T}
   (also \citet{2000ApJ...531..561A,2002PhPl....9.2876A}).
}

\section{Period ratios for standing modes supported by coronal slabs}
\label{sec_Cor_Pratio}

\subsection{Overview of Coronal Slab Dispersion Diagrams}
\label{sec_sub_Cor_DR}

{Let us consider first the coronal case, by which we mean} the ordering $v_{Ae}>v_{A0}>c_0>c_{e}$ holds.
To be specific, we choose $c_0=1$, $v_{A0}=4$ and $c_{e}=0.72$.
Taking the temperature pertaining to TRACE $171$\AA ($\sim 0.96$~MK) as representative,
   one has $c_0 \approx 130$\velunits,
   and hence $v_{A0}=520$\velunits, and $c_e \approx 93$\velunits,
   which are very close to the values derived by~\citet{2001A&A...372L..53N}.
The reference value for the external Alfv\'en speed $v_{Ae}$ is taken to be $2v_{A0}$, corresponding to
   a density contrast $\rho_0/\rho_e = 3.76$ (see Eq.(\ref{eq_rhoe0}) which
   shows that when $c^2 \ll v_A^2$, $\rho_0/\rho_e \approx (v_{Ae}/v_{A0})^2$).
We will also examine the cases where $v_{Ae}/v_{A0}$ is $3$ and $4$,
   which correspond to a $\rho_0/\rho_e$ of $8.41$ and $14.9$, respectively.

Figure~\ref{fig_Cor_DR} presents the phase speeds $\vph$ as a function of longitudinal wavenumber $k$
   for a series of $U_0$.
From top to bottom, Figs.\ref{fig_Cor_DR}a to \ref{fig_Cor_DR}d correspond to an $M_0$ of $0, 0.8, 1.2$
   and $3.2$, respectively.
Here $M_0 = U_0/c_0$ expresses the internal flow speed in units of the internal sound speed.
The dashed (solid) curves are for kink (sausage) modes.
As shown in Fig.\ref{fig_Cor_DR}b, the modes are further labeled by combinations of letters b/f+F/S+K/S, representing
   backward or forward, Fast or Slow, Kink or Sausage.
While ``Fast'' or ``Slow'' is determined by the magnitude of the phase speeds,
   ``Backward'' or ``Forward'' is determined by the sign of the phase speeds when the flow is absent.
The number appended to the letters denotes the order of occurrence.
As such, bFK1 represents the first branch of backward Fast Kink mode, and likewise, fFS2
   the second branch of forward Fast Sausage mode.
On the left of each panel, the characteristic speeds external to the slab are given.
Although nearly indistinguishable, there are two windows where waves are prohibited if their phase speeds are such that
    $c_{Te}^2 < \vph^2 < c_e^2$.
On the right of each panel, characteristic speeds interior to the slab are given.
When $\vph-U_0$ lies in the intervals $(-v_{A0}, -c_0)$,
    $(-c_{T0}, c_{T0})$, or $(c_0, v_{A0})$, the waves are surfaces waves.
Otherwise, they are body ones.
As was pointed out by~\citet{1995SoPh..159..213N} (hereafter NR95), and also evident from Fig.\ref{fig_Cor_DR},
    no surface waves are introduced by the flow in the coronal case.
The rest of the section complements and expands the study by~NR95.

Figure~\ref{fig_Cor_DR} shows that the dispersion diagrams demonstrate a clear dependence on the flow speed.
With increasing $U_0$, there is increasingly less symmetry between the lower and upper half
   of the diagrams, with a perfect symmetry taking place only for a static slab (Fig.\ref{fig_Cor_DR}a).
For the ease of description, let us consider first the slow modes, namely the
   waves with $\vph$ of the order of the sound speeds.
The most obvious variation with varying $U_0$ is that the slow modes are shifted upwards, encompassing
   always the windows $(-c_{0}+U_0, -c_{T0}+U_0)$ and $(c_{T0}+U_0, c_{0}+U_0)$.
This is understandable given that for $kd \rightarrow 0$,
\begin{eqnarray}
 {\vph} \approx U_0 \pm c_{T0} \sqrt{1+ \frac{c_{T0}^4}{c_0^2 v_{A0}^2 h_j^2} k^2 d^2} ,
 \label{eq_Cor_slow_smallk}
\end{eqnarray}
   where
\begin{eqnarray}
   h_j =\left\{\begin{array}{l c}
           (j-1/2)\pi & \mbox{      kink} \\
           j\pi 	  & \mbox{      sausage}
          \end{array} \right.
\label{eq_hj}
\end{eqnarray}
with $j=1, 2,\cdots$.
In the opposite extreme $kd \gg 1$, one may find
\begin{eqnarray}
 \vph \approx  U_0 \pm c_{0} \sqrt{1- \frac{c_{0}^2 g_j^2}{v_{A0}^2 - c_0^2} (k d)^{-2}},
 \label{eq_Cor_slow_bigk}
\end{eqnarray}
where
\begin{eqnarray}
   g_j =\left\{\begin{array}{l c}
           j\pi 	  & \mbox{      kink} \\
           (j-1/2)\pi & \mbox{      sausage}
          \end{array} \right.
\label{eq_gj}
\end{eqnarray}
with $j=1, 2,\cdots$.
In Eqs.(\ref{eq_Cor_slow_smallk}) and (\ref{eq_Cor_slow_bigk}), the plus and minus signs correspond to
   the upper and lower band, respectively.
They also help explain the occurrence of the backward slow waves, labeled BSK and BSS,
   in the upper half of Figs.\ref{fig_Cor_DR}c and \ref{fig_Cor_DR}d:
   even though BSK and BSS have negative phase speeds in a frame comoving with the internal flow,
   in the current frame they have positive speeds as long as $U_0/c_{T0} > 1$ or $M_0 \gtrsim 0.97$.
In view of Eqs.(\ref{eq_hj}) and (\ref{eq_gj}), there actually exists an infinite number of branches of slow modes.
However, in Fig.\ref{fig_Cor_DR} it is neither easy to differentiate these branches,
    or to distinguish between the kink and sausage modes,
    since $c_{0}$ differs only slightly from $c_{T0}$.
Given that the slow modes are nearly nondispersive,
    and that the deviation from unity of the period ratio $P_1/2P_2$
    is due entirely to wave dispersion in the current study,
    we will not examine $P_1/2P_2$ associated with slow modes.

The fast modes are more interesting since they show significant dispersion.
With increasing $U_0$, while the modes in the upper half of the $\vph-k$ plane become less dispersive, the opposite
    is true for the ones in the lower-half:
    the phase speeds of fast modes are bounded in the former by $U_0+v_{A0}$ and $v_{Ae}$,
    and bounded by $U_0-v_{A0}$ and $-v_{Ae}$ in the latter.
It can be readily shown that this behavior results from the fact that for $kd \gg 1$,
\begin{eqnarray}
 \vph \approx  U_0 \pm v_{A0} \sqrt{1+ \frac{v_{A0}^2 h_j^2}{v_{A0}^2 - c_0^2} (k d)^{-2}}.
 \label{eq_Cor_fast_bigk}
\end{eqnarray}
On the other hand, except for the first kink branch, both kink and sausage modes have cutoffs, which are given by
\begin{eqnarray}
 (kd)_c = g_j \Lambda^{\pm},
\label{eq_Cor_fast_cutoff}
\end{eqnarray}
where
\begin{eqnarray*}
 \Lambda^{\pm} = \sqrt{\frac{(c_0^2+v_{A0}^2)[(v_{Ae}\mp U_0)^2-c_{T0}^2]}{[(v_{Ae}\mp U_0)^2-c_{0}^2][(v_{Ae}\mp U_0)^2-v_{A0}^2]}}  .
\end{eqnarray*}
For the first kink branch, an analytic approximation to $\vph$ in the slender slab limit $kd \ll 1$ can be found, which reads
\begin{eqnarray}
 \vph^{\pm} \approx  \pm v_{Ae}\sqrt{1-\eta^{\pm} (kd)^2},
 \label{eq_Cor_FstKnk_smallK}
\end{eqnarray}
    where
\begin{eqnarray}
 \eta^{\pm} = \left(\frac{\rho_0}{\rho_e}\right)^2\frac{[(v_{Ae} \mp U_0)^2-v_{A0}^2]^2 (v_{Ae}^2-c_{e}^2)}{(v_{Ae}^2+c_{e}^2)v_{Ae}^2 (v_{Ae}^2-c_{Te}^2)} ,
\end{eqnarray}
    with $\vph^{+}$ and $\vph^{-}$ representing the upper and lower branches, respectively.

Compared with NR95, the most relevant study to ours,
    we have extended the expressions pertaining to slow waves
    in the slender slab limit (Eq.(16) in NR95)
    to accommodate negative values of phase speeds
    (see Eq.(\ref{eq_Cor_slow_smallk})).
In addition, we have examined the case of wide slabs ($kd \gg 1$, see Eq.(\ref{eq_Cor_slow_bigk}))
    for slow waves.
As for the fast waves, Eq.(\ref{eq_Cor_fast_cutoff}) extends Eq.(13) in NR95
    in the sense that only positive phase speeds are examined in NR95 but the phase speed
    in the present study is allowed to be negative.
Besides, we have presented the expressions of the dispersion behavior in the slender slab limit
    for the first fast kink branch (Eq.(\ref{eq_Cor_FstKnk_smallK})), and obtained new expressions for both kink and sausage waves
    at $kd \gg 1$ (Eq.(\ref{eq_Cor_fast_bigk})).

\subsection{Procedures for Computing Standing Modes}
\label{sec_sub_method}

To begin with, consider a pair of waves corresponding to $(\omega, k_l)$ and $(\omega, k_r)$,
   where $k_r > k_l$ in an algebraic sense.
The combination of the two results in a Lagrangian displacement $\xi_x(x, z; t)$ in the form
\begin{eqnarray}
\xi_x(x, z; t) =
  &&  \mathrm{Re}\left\{\tilde{\xi}_{x,l}(x)\exp\left[-i\left(\omega t - k_l z\right)\right]  \right\} \nonumber \\
  &+& \mathrm{Re}\left\{\tilde{\xi}_{x,r}(x)\exp\left[-i\left(\omega t - k_r z\right)\right] \right\} .
\label{eq_displacement}
\end{eqnarray}
By ``standing'', we require that $\xi_x(\pm d, 0; t) = 0$ and
   $\xi_x(\pm d, L; t) = 0$, where $z=0$ and $L$ denote the slab ends.
Putting $x=d$ and $z=0$ in Eq.(\ref{eq_displacement}), one finds $\tilde{\xi}_{x,l}(d) = - \tilde{\xi}_{x,r}(d)$, and hence
   it makes sense to choose $\tilde{\xi}_{x,l}(d) = A_{\xi}$ to be real.
It then follows that
\begin{eqnarray}
&&  \xi_x(d, z; t) \nonumber \\
&=& A_\xi \left[\cos\left(\omega t - k_l z\right) - \cos\left(\omega t - k_r z\right)\right] \nonumber \\
&=& -2A_\xi\sin\left(\frac{k_r-k_l}{2}z\right)\sin\left(\omega t-\frac{k_l + k_r}{2}z\right) .
\label{eq_stand_displc}
\end{eqnarray}
For $\xi_x(d, L; t)$ to be zero at arbitrary $t$, this requires
\begin{eqnarray}
k_r - k_l =  \frac{2\pi n}{L}, n=1, 2, \cdots
\label{eq_k4stand}
\end{eqnarray}
By convention, $n=1$ corresponds to the fundamental mode, and $n=2$ to its first overtone.

Equation~(\ref{eq_k4stand}) suggests a graphical means to determine the period ratio $P_1/2P_2$
   for a given aspect ratio $d/L$.
This is illustrated by Figure~\ref{fig_Sol_Proc} where we use the results computed for $M_0=0.8$
   as given in Fig.\ref{fig_Cor_DR}b.
Here panels (a) and (b) are for kink and sausage modes, respectively.
First, we convert the $\vph-k$ diagram to an $\omega - k$ diagram.
Only the first and second quadrants are relevant, with the former (latter) coming
   from the curves in the upper (lower) half-plane
   of Fig.\ref{fig_Cor_DR}b.
Note that instead of combinations of positive $k$ with negative $\vph$
   (and hence negative $\omega$), we obtain combinations of negative $k$ with positive $\omega$ by
   reversing the sign of $k$ {in view of symmetry 1}.
This, together with a negative $\vph$, yields a positive $\omega$.
We then compute $P_1/2P_2$ for a given $d/L$ by measuring the width between the intersections of a horizontal line
   with the two curves corresponding to the waves that are to form standing modes.
Take Fig.\ref{fig_Sol_Proc}a for instance.
Let $A$ and $B$ denote the intersections where a horizontal dashed line crosses bFK1 and fFK1, respectively.
The angular frequency $\omega_1$ for the fundamental mode is then
   found by requiring ${AB}_1= 2\pi d/L$, while $\omega_2$ for its first overtone
   is found by requiring that ${AB}_2= 4\pi d/L$.
The period ratio is then $P_1/2P_2 = \omega_2/2\omega_1$.
In addition, from Fig.\ref{fig_Sol_Proc}b one can readily see that
   the minimum allowed aspect ratio, namely $(d/L)_{\mathrm{cutoff}}$, is not determined by the difference between the two
   cutoffs of the two $\omega - k$ curves involved divided by $2\pi$, but larger than that.
We will come back to this point in discussing Fig.\ref{fig_Cor_PrSsg_MA}.
{Furthermore, from symmetry 2 one finds that if $M_0$ is $-0.8$ instead, the $\omega-k$ diagram will be
   a reflection about the vertical axis of what we have for $M_0=0.8$.
It is then clear that the period ratios do not depend on the sign of $U_0$ but only on its magnitude.}

At this point we note that it is impossible for a kink and a sausage wave to combine to form a standing mode.
This is not obvious by examining Eq.(\ref{eq_k4stand}) alone but has to do
    with the symmetric properties of the eigenfunctions.
Let $k_r$ and $k_l$ represent, say, a sausage and kink wave, respectively.
Now that the displacement for kink (sausage) modes is an odd (even) function of $x$,
   in light of Eq.(\ref{eq_stand_displc}) one finds that
\begin{eqnarray*}
   \xi_x(-d, z; t)
&=&  A_\xi \left[\cos\left(\omega t - k_l z\right) + \cos\left(\omega t - k_r z\right)\right]	\\
&=&  2 A_\xi \cos\left(\frac{k_r-k_l}{2}z\right)\cos\left(\omega t-\frac{k_l + k_r}{2}z\right) ,
\end{eqnarray*}
   which does not have a permanent node at $z=0$ or $L$ when $k_l$ and $k_r$ satisfy Eq.(\ref{eq_k4stand}), hence
   violating the requirements for a standing mode.

In {computing the coronal standing modes}, we consider only bFK1 and fFK1 for kink modes,
   and bFS1 and fFS1 for sausage modes.
Higher order branches like bFK2 or bFS2 are not considered since they would form standing modes only
   for relatively thick slabs.
Take fFK2 and bFK2 for instance.
For them to combine to form a standing mode, the slab aspect ratio $d/L$ has to exceed $0.85$.
In the case of fFS1 and bFS2, $d/L$ has to be larger than $0.95$.

Can slow and fast kink (or sausage) waves form a standing mode?
For sausage modes like fSS or bSS to combine with modes like fFS1 or bFS1, the slab has to be very wide.
In the case of fSS and bFS1, the aspect ratio $d/L$ has to exceed $0.72$.
For kink waves, combinations like bFK1 with fSK seem possible but turn out extremely unlikely
   since the slow mode tends to be dominated by the intensity oscillations
   rather than the displacement of the slab, whereas the opposite is true for the kink mode.
For slender slabs $k d \rightarrow 0$, the backward fast kink wave corresponds to $\vph \approx - v_{Ae}$,
    and $n_0 d \rightarrow 0$, and hence
    $X \approx \frac{(\vph - U_0)^2}{(\vph - U_0)^2 - c_0^2}(n_0 d)^2$.
If one neglects the flow and notes that $c_0^2 \ll v_{A0}^2$, then
    one finds that this ratio is
    $\approx (\rho_0/\rho_e-1)(kd)^2$, meaning that fast kink waves are dominated by the displacement rather than the density
    variation.
However, for slow kink waves, by noting that for $kd\rightarrow 0$,
   $(\vph - U_0)^2 \rightarrow c_{T0}^2$, one finds that
   $X \approx -(v_{A0}^2/c_0^2) (n_0 d)\tan(n_0 d)$.
With $n_0 d$ approaching $(j-1/2)\pi$, this will be large, meaning that slow kink waves are dominated by density variations.
The end result is that, if a fast kink wave does combine with a slow one to form a standing mode, a transverse displacement
    on the order of the slab width will cause a relative density variation, and hence a relative intensity variation,
    that may exceed unity!
This makes such combinations unrealistic.

\subsection{Period Ratios for Standing Kink Modes}
\label{sec_sub_Cor_PrKnk}

Figure~\ref{fig_Cor_PrKnk} presents the period ratio $P_1/2P_2$ as a function of the slab aspect
   ratio $d/L$ for standing fast kink modes.
A series of values for the flow speed $U_0$ is adopted, and
   $U_0$ is measured in units of the internal Alfv\'en speed
   $U_0 = M_A v_{A0}$.
Common to all the curves is that they start with unity at small $d/L$, where
   wave dispersion is negligible, decrease with increasing $d/L$ and attain a minimum before
   rising again.
With increasing $U_0$, the deviation of $P_1/2P_2$ from unity is enhanced substantially when compared with a static case
   $M_A=0$.
Take the minima for instance.
While in the static case it reads $0.851$, when $M_A=0.8$ it is significantly reduced
   to $0.657$, amounting to a relative difference of 22.8\%.
One may notice that the former is attained at $d/L=0.137$, while the latter at $d/L=0.139$.
However, this is not to say that for smaller aspect ratios the effects of enhanced flow speed is negligible.
For instance, for $d/L$ as small as $0.03$, while $P_1/2P_2$ reads $0.939$ in the static case,
   it reads $0.85$ when $M_A$ increases to $0.8$.
From this we conclude that, in agreement with \citet{2011A&A...526A..75M}, transverse structuring may contribute
   to the deviation from unity of $P_1/2P_2$ even for thin slabs.
And we note that, in addition to the density contrast which is the sole contributor
   to wave dispersion in \citet{2011A&A...526A..75M}, the velocity shear also plays a significant role.
This has significant bearings on the applications of
   look-up diagrams expressing the period ratio as a function of
   longitudinal density stratification like the one constructed by~\citet{2005ApJ...624L..57A}
   for coronal loops.
While a slab diagram has yet to appear,
   it seems appropriate to say before using this diagram to deduce the longitudinal stratification,
   one had better exclude the wave dispersion brought forth by the transverse structuring.
This is especially necessary for strong transverse density contrasts, and/or in the presence of
   substantial flow shears.

The effect of flow is further shown in Fig.\ref{fig_Cor_PrKnk_MA},
    where (a) the minimal period ratio, $(P_1/2P_2)_{\mathrm{min}}$, as well as (b) its location, $(d/L)_{\mathrm{min}}$,
    are shown as a function of the Alfv\'enic Mach number $M_A$.
In addition to the nominal value of $v_{Ae}/v_{A0}$ being $2$, other ratios of $3$ and $4$
    are also examined and shown in different colors.
Consider Fig.\ref{fig_Cor_PrKnk_MA}a first.
One can see that the effect of flow on the period ratios is significant for
    all the considered $v_{Ae}/v_{A0}$, or equivalently, the density contrast.
Take $v_{Ae}/v_{A0}=4$ for instance.
While the effect of flow is slightly less pronounced than in the case where $v_{Ae}/v_{A0}=2$,
    introducing a flow of $M_A = 0.8$ also rather significantly reduces the period ratio
    from the static value $0.778$ to $0.633$, or by 18.3\%.
Likewise, in the case with $v_{Ae}/v_{A0}=3$, a relative reduction of 20.2\% is found with
    $P_1/2P_2$ being 0.802 for $M_A=0$ but $0.64$ for $M_A=0.8$.
Now consider Fig.\ref{fig_Cor_PrKnk_MA}b.
One may notice that, with increasing density contrast, the aspect ratio at which
    the minimum period ratio is attained decreases.
For a given density contrast, with increasing flow this aspect ratio tends to increase,
    even though the effect is marginal for $v_{Ae}/v_{A0}=2$ and is more pronounced with higher $v_{Ae}/v_{A0}$.
For instance, for $v_{Ae}/v_{A0}=2$, a value of $0.138$ can be quoted for all the $(d/L)_{\mathrm{min}}$ obtained,
    whereas for $v_{Ae}/v_{A0}=4$, $(d/L)_{\mathrm{min}}$ increases substantially from 0.057 in the static case
    to $0.106$ when $M_A=0.8$.
Furthermore, the overall tendency for both $(P_1/2P_2)_{\mathrm{min}}$ and $(d/L)_{\mathrm{min}}$ to decrease with
    increasing density contrast agrees closely with Fig.10 in~\citet{2011A&A...526A..75M}.
However, while the minimum in the static case cannot go below
    $\sqrt{2}/2=0.707$ as determined analytically
    for an Epstein density profile and numerically for a step function
    in~\citet{2011A&A...526A..75M},
    it is not subject to this constraint in a flowing case.

\subsection{Period Ratios for Standing Sausage Modes}
\label{sec_sub_pratio_sausage}

Now consider standing sausage modes.
Figure~\ref{fig_Cor_PrSsg} presents the period ratio $P_1/2P_2$ as a function of aspect ratio $d/L$
   for both the cases where $v_{Ae}/v_{A0}=2$ (solid curves)
   and $v_{Ae}/v_{A0}=4$ (dashed curves).
A series of $U_0$, once again expressed in terms of the Alfv\'enic Mach number $M_A$,
   is adopted and indicated by different colors.
For the case $v_{Ae}/v_{A0}=2$, one can see that except for the results corresponding to the
   first three values of $M_A$, the curves are absent
   since the range of $d/L$ in which standing sausage modes are allowed
   is outside the presented $d/L$ range.
For the three curves that are present, while one can hardly tell them apart when
   their overlapping parts are concerned,
   it is evident that they start with considerably different values of the allowed $d/L$.
With $M_A$ increasing from $0$ to $0.1$ to $0.2$, this cutoff aspect ratio $(d/L)_{\mathrm{cutoff}}$
   increases from $0.3$ to $0.33$ to $0.39$.
This means that at a given $v_{Ae}/v_{A0}$, relative to the static case, slabs with flow can support
   standing sausage modes only when they are sufficiently wide.
The same is also true for other choices of $v_{Ae}/v_{A0}$: when $v_{Ae}/v_{A0}=4$, while $(d/L)_{\mathrm{cutoff}}$
   reads $0.134$ for the static case, it reads $0.157$ when $M_A=0.2$ and $0.232$ when $M_A=0.4$.
In this case, the effect of flow on the period ratio turns out to be more pronounced than in the case $v_{Ae}/v_{A0}=2$.
At $d/L=0.35$, the case with $M_A=0.5$ yields a $P_1/2P_2$ of $0.654$, which is 5.2\% lower than the value $0.69$
   obtained in the static case.
If comparing the solid and dashed curves, one can see that the period ratio as well as
   the cutoff aspect ratio depend rather
   sensitively on $v_{Ae}/v_{A0}$, consistent with
    \citet{2011A&A...526A..75M}.
Here we have shown the new result that, for standing sausage modes,
   the flows are important in determining the ranges of aspect ratios
   where standing modes are allowed.
For the considered density contrasts the relative changes
   in the period ratio due to flows are not as significant as in the standing kink cases, though.

Before proceeding, we note in passing that due to the strong dispersion accompanying sausage waves,
   it is natural to expect the period ratio $P_1/2P_2$ to deviate from $1$.
As demonstrated by~\citet{2011A&A...526A..75M}, for static coronal slabs, $P_1/2P_2$ may reach values as low as $1/2$ when
   the density contrast increases towards infinity.
How low $P_1/2P_2$ is allowed to go for coronal loops with finite aspect ratios will be explored in a future
   study that extends the slab computations presented here and in~\citet{2011A&A...526A..75M}.
Regardless, it is safe to say that $P_1/2P_2$ will deviate in general from $1$ even in the absence
   of any longitudinal structuring since sausage waves are also strongly dispersive in a cylindrical geometry.
This implies that practices employing standing sausage modes to infer longitudinal density scale heights may
   be problematic~\citep{2008MNRAS.388.1899S}.

Figure~\ref{fig_Cor_PrSsg_MA} further examines the effects of flow speed on
   the cutoff aspect ratio $(d/L)_{\mathrm{cutoff}}$ pertinent to the standing sausage modes.
Here three cases with different values of $v_{Ae}/v_{A0}$ are examined, and
   presented with different colors.
Besides, the dashed curves represent
   the distribution with $M_A$ of the lower limit of $(d/L)_{\mathrm{cutoff}}$
   given by
   $(\Lambda^{+}+\Lambda^{-}) g_1/2\pi = (\Lambda^{+}+\Lambda^{-})/4$ (see Eq.(\ref{eq_Cor_fast_cutoff})).
It is evident that with increasing $M_A$,
   the deviation of the computed values from this limit increases substantially.
Note that because $c_0^2 \approx c_{T0}^2 \ll v_{A0}^2$, this limit can be satisfactorily approximated by
\begin{eqnarray*}
\frac{1}{4}\left(1/\sqrt{\left(\frac{v_{Ae}}{v_{A0}}-M_A\right)^2-1}
                +1/\sqrt{\left(\frac{v_{Ae}}{v_{A0}}+M_A\right)^2-1}\right),
\end{eqnarray*}
    which has only a modest $M_A$ dependence in the parameter range considered.
The fact that the computed
    cutoff aspect ratios have a much stronger $M_A$ dependence results from an increasingly strong
    asymmetry in the $\omega - k$ diagram, as discussed regarding Fig.\ref{fig_Sol_Proc}b.
From an observational viewpoint, the curves presented in Fig.\ref{fig_Cor_PrSsg_MA}
    can be tested in a statistical sense if there is a large set of slab-supported
    standing sausage oscillations.
Sorting the observations into groups with similar density contrasts,
    and in any group constructing a parameter space comprising the aspect ratio
    as well as the Alfv\'{e}nic Mach number,
    then one expects to see a ridge across which detections of such standing modes
    become drastically less frequent.

A few words are in order to justify our examination of standing sausage modes
  for slabs with a finite width.
Indeed, loops in the form of bright structures appear to be thin in most UltraViolet images of the solar corona,
  meaning that thin-tube limits are often valid in many seismological applications.
However, coronal loops with finite aspect ratios may exist.
For instance, the flaring loop examined 
  in~\citet{2003A&A...412L...7N} is $25$~Mm long, and $6$~Mm in diameter, resulting in an aspect ratio
  of $0.12$ using our definition.
That loop was shown to be wide enough to support a global sausage mode.
{However, one may ask how typical these ``fat'' structures are, given that their observations seem to be rather sparse.
With this caveat in mind}, the results found here can {nevertheless} serve as a reference for our photospheric computations to compare with.

\section{Period ratios for standing modes supported by photospheric slabs}
\label{sec_Pho_Pratio}

\subsection{Overview of Photospheric Slab Dispersion Diagrams}
\label{sec_sub_Pho_DR}

Now move on to the photospheric case, by which we mean the ordering $v_{A0}>c_{e}>c_0>v_{Ae}$ holds.
Following NR95 and~\citet{2003SoPh..217..199T}, we deal only with an isolated slab immersed
   in an unmagnetized atmosphere:
   $c_0 = 1, v_{A0} = 1.5, c_e = 1.2, v_{Ae}=0$ (and hence $c_{T0} = 0.83, c_{Te} = 0, \rho_e/\rho_0 = 2$).
Putting some absolute value for $c_0$ as $8$\velunits, one finds $v_{A0} = 12$,
   $c_e = 9.6$\velunits, which agree with the photospheric values adopted by NR95,
   and fall in category (ii) in~\citet{1990ApJ...348..346E}. 

Figure~\ref{fig_Pho_DR} presents, in a format similar to Fig.\ref{fig_Cor_DR},
   the phase speeds $\vph$ as
   a function of longitudinal wavenumber $k$
   for a series of $U_0$.
Here again $U_0$ is measured in units of the internal sound speed $M_0 = U_0/c_0$,
   and kink (sausage) modes are given by the dashed (solid) curves.
Labeling different waves is slightly more complicated than in the coronal case,
   since now in addition to the body waves,
   surface waves are also permitted and
   can be further grouped into fast and slow ones~\citep{1990ApJ...348..346E}.
However, given that body waves cannot combine with surface ones
   to form standing modes (see section~\ref{sec_sub_comp_pho_pratios}),
   and propagating body waves show only a modest dispersion,
   the flow effect turns out to be marginal for pure standing body modes.
Therefore we leave them out, and label surface waves only, as indicated by Fig.\ref{fig_Pho_DR}a.
With the convention ``b/f+F/S+K/S'' we have, for instance,
   fFS for forward fast sausage waves, and bSK for backward slow kink waves.

Figure~\ref{fig_Pho_DR} indicates that the dispersion diagrams possess
    a stronger dependence on the internal flow
    than in the coronal case, evidenced by the fact that some propagating windows may disappear.
Take the body waves for example, for which two bands exist and are shifted upwards with increasing $U_0$ .
For forward (backward) ones, the phase speeds $\vph$ start with $c_{T0}+U_0$ ($-c_{T0}+U_0$) at $kd \ll 1$
    in the same manner as described by Eq.(\ref{eq_Cor_slow_smallk}), for both kink and sausage waves alike.
With increasing $kd$, $\vph$ approaches $\pm c_0 + U_0$ in the way
    given by Eq.(\ref{eq_Cor_slow_bigk}), the only exception is the first branch of kink waves:
    $\vph$ may exceed $c_{0}+U_0$ (the first dashed curve from top in Fig.\ref{fig_Pho_DR}a)
    or fall below $-c_{0}+U_0$ (the first dashed curves from bottom in all panels),
    thereby rendering these body waves a surface one asymptotically.
This phenomenon, termed mode crossing, was also shown in~\citet{2003SoPh..217..199T}.
When $c_e - c_{T0} > U_0 > c_e - c_0$ (Fig.\ref{fig_Pho_DR}c),
    the upper window is bounded from above by $c_e$ instead
    of $c_0+U_0$, therefore providing a maximum allowed longitudinal wavenumber
    beyond which forward slow body kink waves are no longer trapped. 
When $U_0 > c_e - c_{T0}$ (Fig.\ref{fig_Pho_DR}d), this upper window disappears altogether.

Surface waves are examined hereafter.
Consider first the fast ones.
One finds that for $kd \ll 1$, only sausage waves are allowed, kink waves appear only at
    relatively large $kd$ as a continuation of the body kink waves, i.e., via mode crossing.
This is readily understandable given that with $kd \rightarrow 0$ and $\vph^2 \rightarrow c_e^2$,
    if $m_0 d \rightarrow 0$ then
    the DR for kink waves (Eq.(\ref{eq_DR_surface})) does not allow any solution since its LHS
    approaches infinity.
For the fast sausage wave, one has
\begin{eqnarray}
 {\vph} \approx \pm c_e \sqrt{1- \frac{\rho_e^2}{\rho_0^2}
     \left(\frac{c_e^2}{v_{A0}^2}\frac{1-\bar{c}_{e,\pm}^2/c_{0}^2}{1-\bar{c}_{e,\pm}^2/c_{T0}^2}\right)^2 k^2 d^2} ,
 \label{eq_Pho_surf_FstSsg_smallk}
\end{eqnarray}
where $\bar{c}_{e,\pm} = \pm c_e -U_0$.
Now consider the slow surface waves.
One can readily see that for sausage waves $\vph$ starts with $\pm c_{T0}+U_0$,
   and for small $k$,
\begin{eqnarray}
 {\vph} \approx \pm c_{T0}
   \sqrt{1- \frac{\rho_e}{\rho_0 \sqrt{1-{c}_{T0,\pm}^2/c_e^2}} \frac{{c}_{T0,\pm}^2 c_{T0}^2}{v_{A0}^4} k d} ,
 \label{eq_Pho_surf_SlwSsg_smallk}
\end{eqnarray}
where ${c}_{T0,\pm} = \pm c_{T0} +U_0$.
At sufficiently strong internal flow with $U_0 > c_e-c_{T0}$, the upper branch of slow sausage waves
   starts with $c_e$ instead, and the behavior of $\vph$ at small $k$ is identical to Eq.(\ref{eq_Pho_surf_FstSsg_smallk})
   with the plus sign.
On the other hand, for kink waves, $\vph$ at small $k$ can be approximated by
\begin{eqnarray}
 {\vph} \approx \pm \sqrt{\frac{\rho_0}{\rho_e}(v_{A0}^2 - U_0^2) kd} .
 \label{eq_Pho_surf_SlwKnk_smallk}
\end{eqnarray}

The study so far conducted in this section differs substantially from NR95 (their section 3.3), even though
 both studies are for isolated photospheric slabs.
NR95 examined the consequences of a variable $U_e$ while keeping $U_0=0$, as opposed to our study where
    $U_e=0$ but $U_0$ is allowed to vary.
A situation similar to our choice was examined in~\citet{2003SoPh..217..199T} albeit a cylindrical geometry
    is considered there.

\subsection{Computing Standing Modes}
\label{sec_sub_comp_pho_pratios}

As have been discussed, when constructing standing modes, we do not employ pure body waves in view of their modest dispersion.
Once again, by saying it is possible or impossible for a pair of propagating waves to form a standing mode, we mean
   the standing mode hence constructed corresponds
   to a realistic density fluctuation $X$ in the slender slab limit $kd \rightarrow 0$
   provided that the waves have no cutoff longitudinal wavenumber.

Consider kink waves for $kd \rightarrow 0$ first.
To this end, let us recall that fast kink surface waves do not exist, whereas 
   for slow kink surface waves one finds that $X \approx U_0^2/(c_0^2-U_0^2) (m_0 d)^2$ by noting 
   that $\vph \rightarrow 0$ and $m_0 d \rightarrow 0$.
On the other hand, the kink body waves in the limit $kd \rightarrow 0$
   is proportional to $(n_0 d) \tan(n_0 d)$ which approaches infinity given that $n_0 d \rightarrow (j-1/2)\pi$.
This means that standing kink modes can result from a combination of either pure body or pure surface waves, 
   but not a mixture of both.
Therefore we are left with only one task of examining pure kink surface modes.

Standing sausage modes can be considered in a similar manner.
Once again, they cannot be formed by combining a pair of body and surface waves.
This is because for body waves, $X$ is proportional to $(n_0 d) \cot (n_0 d)$ and approaches infinity when $kd$ approaches zero
    given that $n_0 d \rightarrow j\pi$.
For surface waves, however, one finds that at $kd\ll 1$,  for fast waves $X \approx c_e^2/(c_0^2-c_e^2)$,
    whereas for slow waves    
\begin{eqnarray}
X \approx
  \left\{\begin{array}{c c}
           \frac{v_{A0}^2}{c_0^2} ,                         & \mbox{if } U_0 \le c_e - c_{T0} \\ 
           \frac{(c_e-U_0)^2}{c_0^2-(c_e-U_0)^2} .          & \mbox{if } U_0 \ge c_e - c_{T0} 
         \end{array}
  \right.
\end{eqnarray}
Note that the latter holds since $\vph \rightarrow c_e$ when $U_0 \ge c_e - c_{T0}$.
Nonetheless, different from the coronal case, now in the surface category fast and slow waves are allowed to form standing sausage modes.

\subsection{Period Ratios for Standing Kink Modes}
\label{sec_sub_Pho_PrKnk}

Figure~\ref{fig_Pho_PrKnk} presents the period ratio $P_1/2P_2$ as a function of the slab aspect
   ratio $d/L$ for standing kink surface modes.
A series of values for the flow speed $U_0$ is adopted and given by different colors.
Common to all curves is that they start with $\sim 1.4$ at small $d/L$ and
   decrease monotonically with $d/L$ towards unity.
The behavior for slender slabs is in stark contrast to the coronal
   case where for extremely thin slabs
   $d/L \rightarrow 0$ the period ratio approaches unity no matter how strong $U_0$ is.
What happens in the photospheric case is, irrespective of $U_0$,
   $P_1/2P_2$ differs substantially from unity even for zero-width slabs!
This is understandable given that for small $kd$ the surface kink waves have phase speeds
   $\vph \propto \sqrt{kd}$, leading to 
   that the angular frequency $\omega = \vph k$ is proportional to $k^{3/2}$.
Note that, as illustrated by Fig.\ref{fig_Sol_Proc}, the period ratios are computed via
   $P_1/2P_2 = \omega_2/2\omega_1$, where $\omega_1$ and $\omega_2$ correspond to the angular frequencies
   found at the separation $AB$ being $2\pi d/L$ and twice that, respectively.
This means, at small $d/L$, $P_1/2P_2 \approx 2^{3/2}/2 = \sqrt{2}=1.414$.

One can see that the effect of flow on $P_1/2P_2$ is substantial for $d/L$ only in the range between $0.03$ and $0.27$.
Take $d/L=0.1$ for example. 
While $P_1/2P_2 = 1.24$ in the static case, it reads $1.1$ when $M_0 = 0.4$, amounting to a reduction of $11.6\%$.
When $d/L\lesssim 0.03$, the $k^{1/2}$ dependence of phase speed dominates, leading to little difference between different curves.
To the contrary, when $d/L \gtrsim 0.27$, all curves show little deviation from unity since the wave dispersion is only modest
   (see Fig.\ref{fig_Pho_DR}).

\subsection{Period Ratios for Standing Sausage Modes}
\label{sec_sub_Pho_PrSsg}

Consider now standing sausage surface modes.
Figure~\ref{fig_Pho_PrSsg} presents the period ratio $P_1/2P_2$ as a function of aspect ratio $d/L$
   for a series of $U_0$.
Distinct from the coronal case, now there is no cutoff in $d/L$ below which no standing modes can be constructed.
In addition, the combinations of fast and slow waves are now allowed, resulting 
   in four pairs of possibilities.
Note that instead of four, there are only two curves in Figs.\ref{fig_Pho_PrSsg}b and \ref{fig_Pho_PrSsg}d,
   since forward fast waves are absent in the rest of the cases considered. 
In the parameter range we explored, the strongest $U_0$ dependence of $P_1/2P_2$ 
   happens in the case where backward fast waves combine with forward slow waves, presented in Fig.\ref{fig_Pho_PrSsg}c.
The trend of $P_1/2P_2$ with changing $d/L$ is the same for all curves in this case,
    decreasing from unity at small $d/L$
    to some minimum and then increasing toward unity again.
Even in this case, the flow effect is only modest, as evidenced by the fact that for $M_0=0.4$,
    the minimum $P_1/2P_2$ attains is $0.909$, only slightly smaller than $0.965$ attained in the static case.  

{Directly measuring sausage oscillations with imaging instruments was not possible until very recently~\citep{2011ApJ...729L..18M}
  (hereafter MEJM, also see \citeauthor{2008IAUS..247..351D}~\citeyear{2008IAUS..247..351D}, \citeauthor{2012NatCom..3.1315M}~\citeyear{2012NatCom..3.1315M}).
By using the Rapid Oscillations in the Solar Atmosphere (ROSA) instrument situated at the Dunn Solar Telescope,
  MEJM identified sausage modes in magnetic pores by examining the phase relations between 
  the pore size and its intensity.
A number of periods were present, and the authors suggested that some of the values
  may be attributed to the overtones of the fundamental mode, with the standing mode set up by reflections between
  the photosphere and transition region.
Let $P_n$ ($n=2, 3, ..$) denote the period of the overtones with $P_1$ standing for the fundamental.
For now, let us consider only the period $281\pm 18$~s found in the intrinsic mode function (IMF) c4 therein.
Choosing $P_1$ to be $550$~s pertinent to fast modes, this value
  would correspond to a $P_1/nP_n$ in the range $[0.92, 1.05]$ ($[0.61, 0.70]$)
  provided it corresponds to $n=2$ ($n=3$).
Note that the aspect ratio of the structure in question is $d/L\approx 0.3$, since the average pore size
  is $2d \approx 1.16 \times 10^3$~km (corresponding to an area of $1.36\times 10^6$~km$^2$) and $L=2\times 10^3$~km.
At this aspect ratio, the computed $P_1/2P_2$ as given in our Fig.\ref{fig_Pho_PrSsg} indeed lies in the range
   that corresponds to $n=2$, reinforcing the suggestion by MEJM that it is the first overtone.
Nevertheless, choosing $P_1$ to be $660$~s pertinent to slow modes, 
   this period would correspond to a $P_1/nP_n$ in the range $[1.1, 1.26]$ ($[0.74, 0.84]$)
   provided it corresponds to $n=2$ ($n=3$).
While the former seems inappropriate, the latter is plausible given that the dispersion
  introduced by transverse structuring in density and magnetic field strength, along with the possible role
  of velocity shear, reduces $P_1/n P_n$ from unity for photospheric sausage modes.
While this comparison is admittedly inconclusive, the point we want to
   make is that the transverse structuring does broaden the range of
   the period ratios, thereby offering more possibilities in interpreting the measured oscillation periods.
Instead of proceeding further with this discussion, let us note that the slab model may be seen as
   only a first approximation
   of solar atmospheric structures, and the oscillations measured by ROSA may correspond to propagating rather than standing waves,
   as was pointed out by MEJM.
}

\section{Summary and Concluding Remarks}
\label{sec_summary}

The present study has been motivated by two strands of research in the field of { solar magneto-seismology}.
On the one hand, exploiting the measured multiple periods is playing
    an increasingly important role in deducing the longitudinal structuring
   of {solar atmospheric} structures~\citep{2009SSRv..149....3A,2009SSRv..149..199R}
On the other hand, it was recently realized that significant flows in coronal structures reaching the Alfv\'enic regime
    can bring significant revisions to the seismologically deduced
    physical parameters such as the coronal loop magnetic field strength~\citep{2011ApJ...729L..22T}.
To complement and expand the only theoretical study on multiple periodicities in coronal slabs~\citep{2011A&A...526A..75M},
    we examine the consequences the flows may have.
To this end, we numerically solve the dispersion relations for waves supported by slabs incorporating flows,
    and devise a graphical method to construct standing kink and sausage modes.

For coronal slabs, we find that the internal flow has significant effects on the standing modes,
    for the fast kink and sausage ones alike.
For the kink ones, they may reduce the period ratio by up to 23\% compared with the static case,
    and the minimum allowed period ratio may fall below the analytically derived lower limit in the static case.
In particular, the reduction due to a finite flow may be significant even for thin slabs.
For the sausage modes, while introducing the flow leads to a reduction in the period ratio that is typically $\lesssim 5$\%
    relative to the static case,
    it has significant effects on the threshold aspect ratio only above which standing sausage modes can be supported.
At a given density contrast, this threshold may exceed its static counterpart by several times for the
    parameters considered, thereby limiting the detectability of standing sausage modes to even wider slabs.

For the isolated photospheric slabs, we find that the internal flow has only a marginal effect
    on the period ratios $P_1/2P_2$ for the surface modes,
    and even less for body modes.
That said, standing modes supported in this case are distinct from the coronal case in a number of aspects.
First, standing sausage modes may be supported by slabs with arbitrary aspect ratios and do not suffer from a cutoff $d/L$ any longer.
Second, for standing kink surface modes, $P_1/2P_2$ deviates from unity even for a zero-width slab, which originates from
    the $k^{1/2}$ dependence of phase speeds of slow kink surface waves in the slender slab limit.

The present study can be extended in several ways.
First, to be able to directly compare with coronal loop oscillations, so far the most observed and the best understood,
    we need to turn to the cylindrical geometry and examine how introducing the flow
    affects the dispersion behavior and consequently period ratios and cutoff aspect ratios.
Second, allowing the tube parameters to be time-dependent has been found important as far as the seismological applications
    of the period ratio are concerned~\citep{2009A&A...502..315M}.
This is expected to be also true when the flow speeds are dynamic rather than time-stationary.
Third, concerning the period ratios, one important contributor to its deviation from unity that has received much attention
    is the longitudinal structuring~\citep{2009SSRv..149....3A}, which has yet to be considered
    in examining the slab period ratios.
Nonetheless, the main results presented here, which demonstrated the influences of the introduced flow
    on the standing kink and sausage modes, point to the need for further investigations into such influences.

{     
Before closing, let us recall that solar magneto-seismology (SMS) has extended well beyond the territory originally taken up
    by ``coronal seismology''.
One thing to note in particular, as advocated by~\citet{2011ApJ...729L..18M}, is to seriously consider applying SMS to photospheric
    structures because there exist ample observations of oscillatory motions on the one hand,
    relevant theories are available on the other.
Strictly speaking, our results on the sausage modes are more applicable to the lower parts of the solar atmosphere.
In this sense, our computations strengthen the notion raised by~\citet{2011ApJ...729L..18M} that photospheric studies in the context of
    SMS offer equally rich physics as their coronal counterparts.
}

\acknowledgements
We thank the anonymous referee for the constructive comments and suggestions which helped 
   improve this paper substantially.
This research is supported by the National Natural Science Foundation of China (40904047, 41174154, and 41274176), 
  the Ministry of Education of China
  (20110131110058 and NCET-11-0305), and by the Provincial Natural Science Foundation of Shandong via Grant JQ201212.

\begin{figure}
\centering
\epsscale{.5}
\plotone{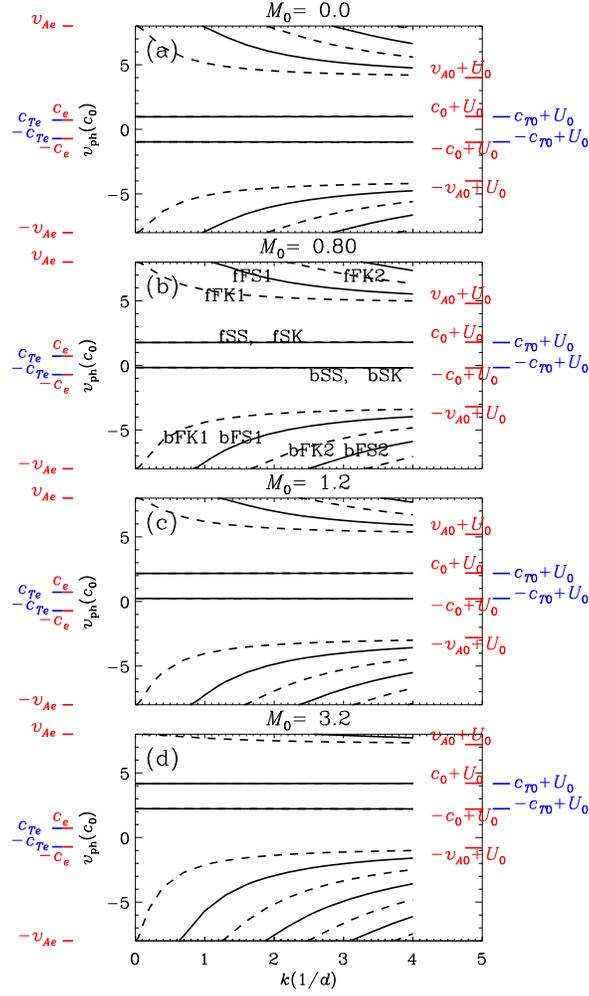}
\caption{Phase speeds $\vph$ as a function of longitudinal wavenumber $k$ for a series of internal flow $U_0$.
Expressing $U_0$ in units of the internal sound speed $U_0 = M_0 c_0$,
   panels (a) to (d) correspond to an $M_0$ of $0, 0.8, 1.2$ and $3.2$, respectively.
On the left (right) of each panel, the characteristic speeds external (interior) to the slab are given
   by the horizontal bars.
Kink and sausage modes are presented by the dashed and solid curves, respectively.
They are further labeled, as shown in panel (b), using combinations of letters b/f+F/S+K/S, representing
   Backward or Forward, Fast or Slow, Kink or Sausage.
The number appended to the letters denote the order of occurrence.
Hence, bFK1 represents the first branch of backward Fast Kink mode.
Moreover, here $v_{Ae}=8$, $c_e=0.72$, $c_{Te}=0.719$, while $v_{A0}=4$, $c_0 = 1$, and $c_{T0}=0.97$.
}
 \label{fig_Cor_DR}
\end{figure}

\begin{figure}
\centering
\epsscale{.7}
\plotone{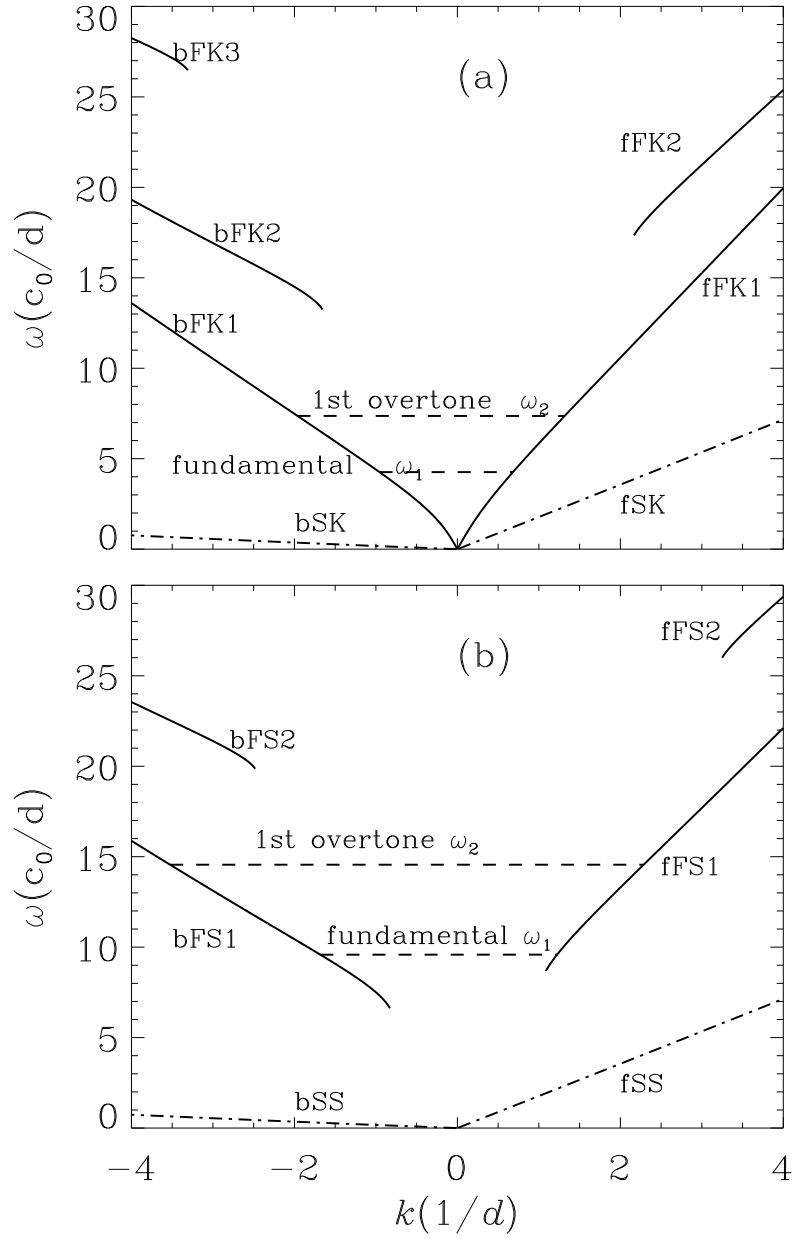}
 \caption{Determination of the period ratio with $\omega-k$ dispersion diagrams for (a) standing kink and (b)
   standing sausage modes.
Different branches labeled with combinations of letters and numbers follow the
   same convention as in Fig.\ref{fig_Cor_DR}.
}
 \label{fig_Sol_Proc}
\end{figure}

\begin{figure}
\centering
\epsscale{1.}
\plotone{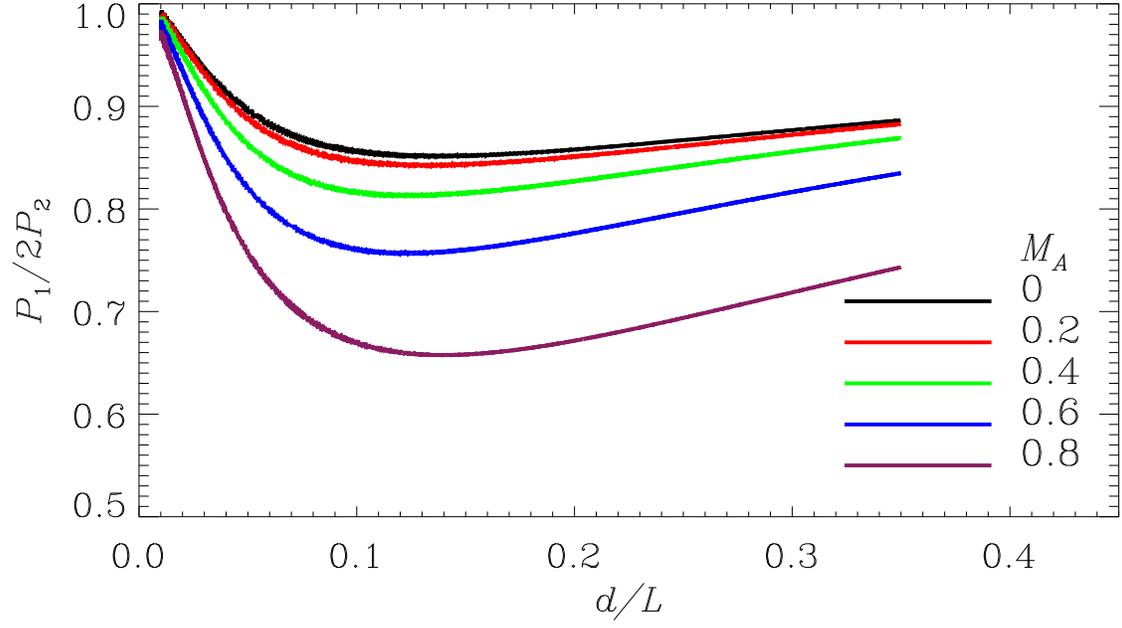}
 \caption{Period ratio $P_1/2P_2$ as a function of the slab aspect
   ratio $d/L$ for standing fast kink modes.
Curves with different colors represent results computed for different values of the
   flow speed $U_0$ measured in units of the internal Alfv\'{e}n speed
   $U_0 = M_A v_{A0}$.
}
 \label{fig_Cor_PrKnk}
\end{figure}

\begin{figure}
\centering
\epsscale{1.}
\plotone{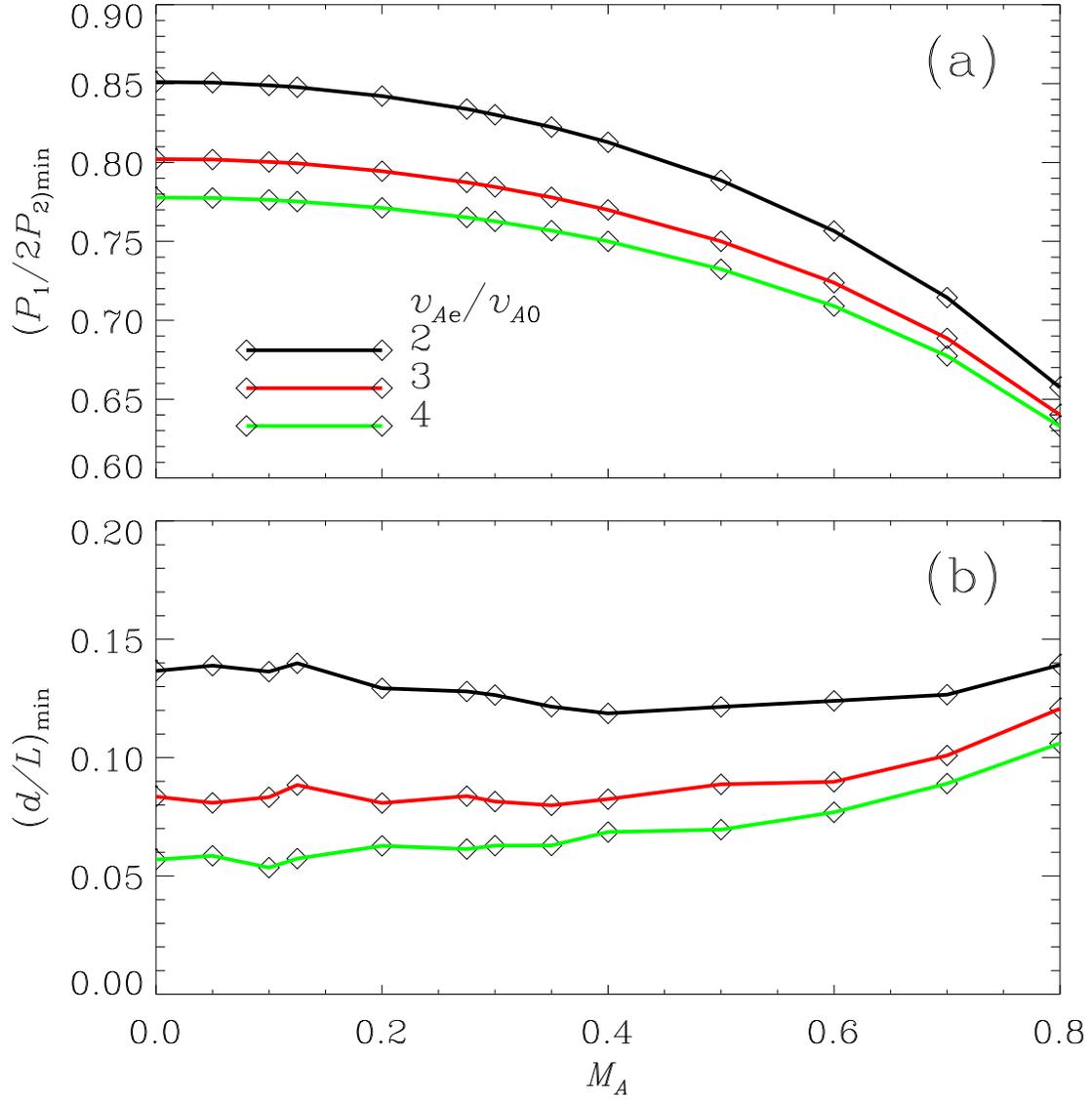}
 \caption{Effects of flow speed on (a) the minimal period ratio, $(P_1/2P_2)_{\mathrm{min}}$, and (b) the aspect ratio
    at which the minimum is attained, $(d/L)_{\mathrm{min}}$.
Here both $(P_1/2P_2)_{\mathrm{min}}$ and $(d/L)_{\mathrm{min}}$
    are displayed as a function of the Alfv\'enic Mach number $M_A$.
}
 \label{fig_Cor_PrKnk_MA}
\end{figure}

\begin{figure}
\centering
\epsscale{1.}
\plotone{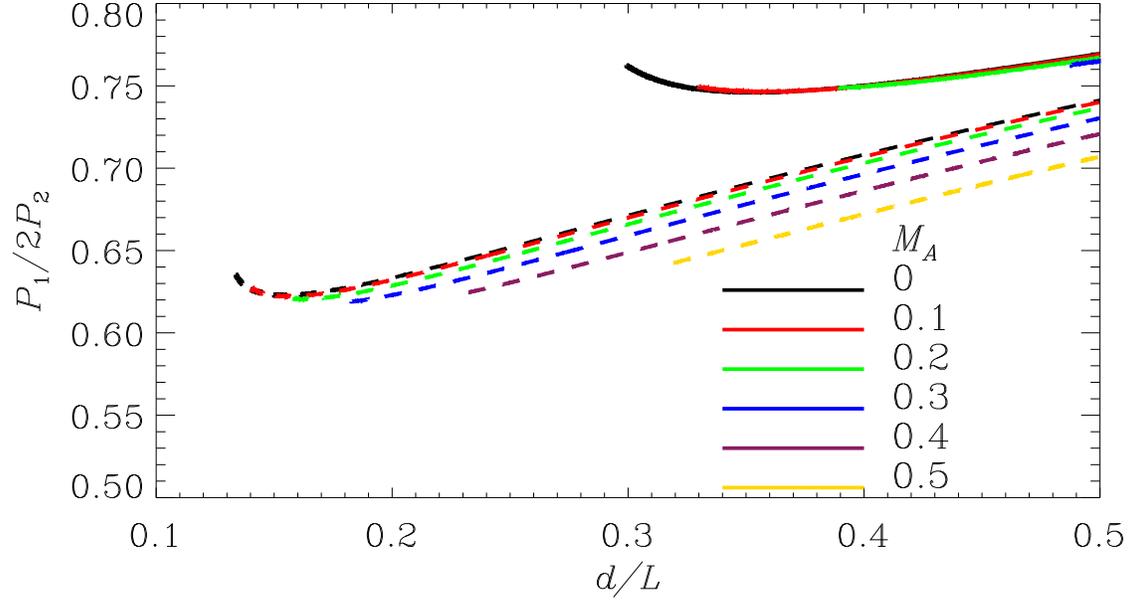}
 \caption{Period ratio $P_1/2P_2$ as a function of the slab aspect
   ratio $d/L$ for standing sausage modes.
The solid and dashed curves are for the cases where $v_{Ae}/v_{A0}=2$ and $4$, respectively.
Curves with different colors represent results computed for different values of the
   flow speed $U_0$ measured in units of the internal Alfv\'en speed
   $U_0 = M_A v_{A0}$.
}
 \label{fig_Cor_PrSsg}
 \end{figure}

\begin{figure}
\centering
\epsscale{1.}
\plotone{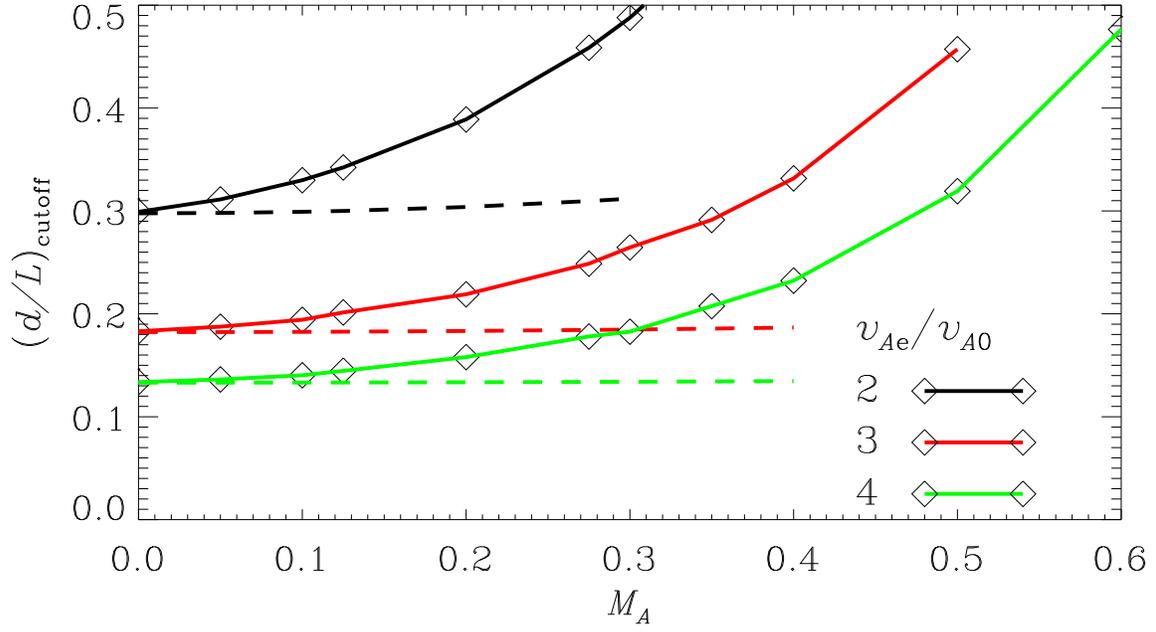}
 \caption{Effects of flow speed on the cutoff aspect ratio, $(d/L)_{\mathrm{cutoff}}$, the lowest allowed
    for standing sausage modes to occur.
Here $M_A$ measures the internal flow speed $U_0$ in units of the internal Alfv\'en speed,
    and different colors are for different ratios of $v_{Ae}/v_{A0}$, ranging from $2$ to $4$.
The dashed curves are the lower limit for $(d/L)_{\mathrm{cutoff}}$, accurate for the static case (please see text
    for details).
}
 \label{fig_Cor_PrSsg_MA}
\end{figure}

\begin{figure}
\centering
\epsscale{.5}
\plotone{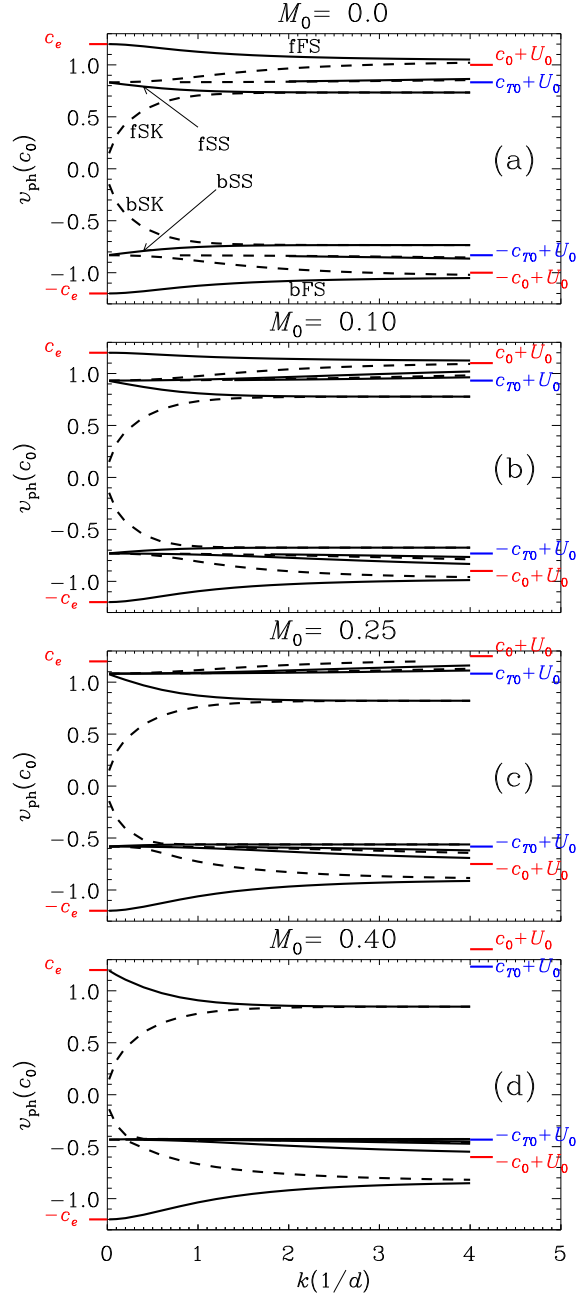}
\caption{Similar to Fig.\ref{fig_Cor_DR} but for an isolated photospheric slab ($v_{Ae} =0$).
The characteristic speeds are $c_e=1.2 c_0, v_{A0}=1.5 c_0$ and $c_{T0}=0.83 c_0$.
Panels (a) to (d) correspond to an $M_0$ of $0, 0.1, 0.25$ and $0.4$, respectively, where
   $M_0 = U_0/c_0$ measures the internal flow in units of the internal sound speed.
Body waves, namely those with phase speeds between $\pm c_{T0}+U_0$ and $\pm c_{0}+U_0$, are not labeled.
Surface waves are labeled with combinations of letters b/f+F/S+K/S, representing
   backward or forward, Fast or Slow, Kink or Sausage.
}
 \label{fig_Pho_DR}
\end{figure}

\begin{figure}
\centering
\epsscale{.8}
\plotone{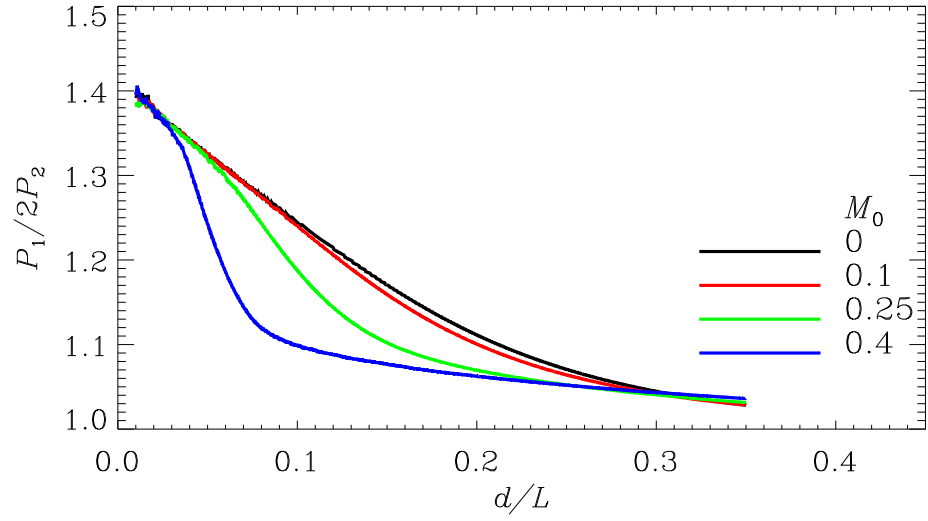}
 \caption{Period ratio $P_1/2P_2$ as a function of the slab aspect
   ratio $d/L$ for standing slow surface kink modes.
Curves with different colors represent results
   computed for different values of the
   flow speed $U_0$ measured in units of the internal sound speed
   $U_0 = M_0 c_{0}$.
}
 \label{fig_Pho_PrKnk}
\end{figure}

\begin{figure}
\centering
\epsscale{.6}
\plotone{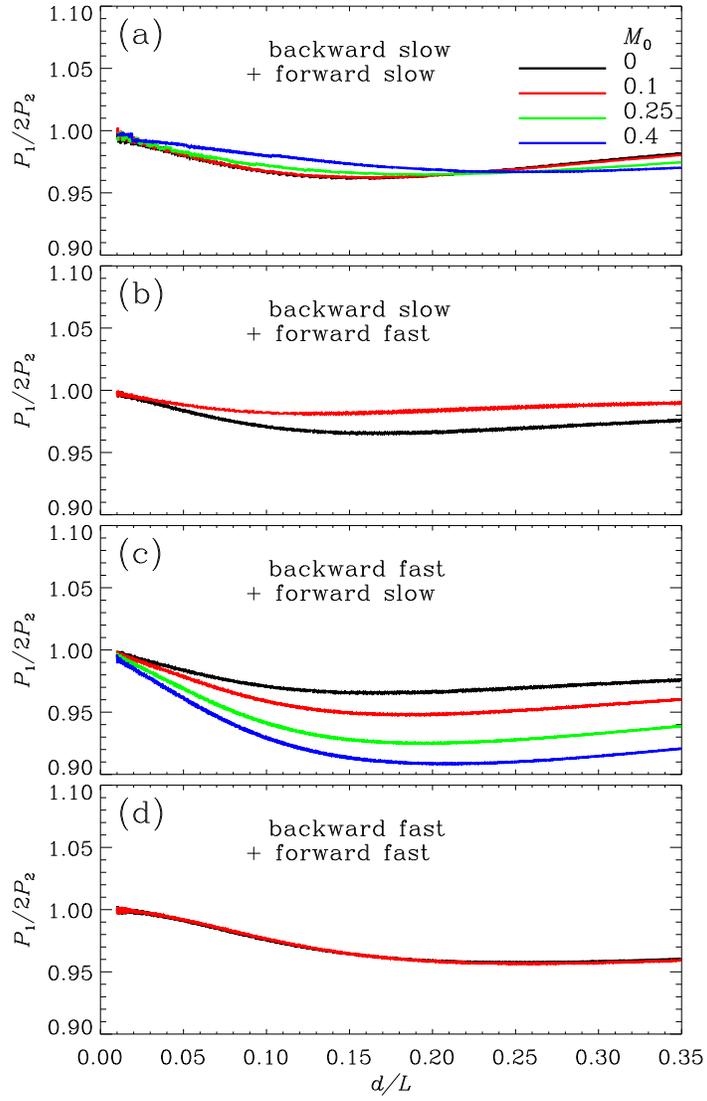}
 \caption{Period ratio $P_1/2P_2$ as a function of the slab aspect
   ratio $d/L$ for standing surface sausage modes.
Presented are combinations of (a) backward slow + forward slow,
   (b) backward slow + forward fast,
   (c) backward fast + forward slow,
   and (d) backward fast + forward fast waves.
Curves with different colors represent results
   computed for different values of the
   flow speed $U_0$ measured in units of the internal sound speed
   $U_0 = M_0 c_{0}$.
}
 \label{fig_Pho_PrSsg}
 \end{figure}

\begin{figure}
\centering
\epsscale{.8}
\plotone{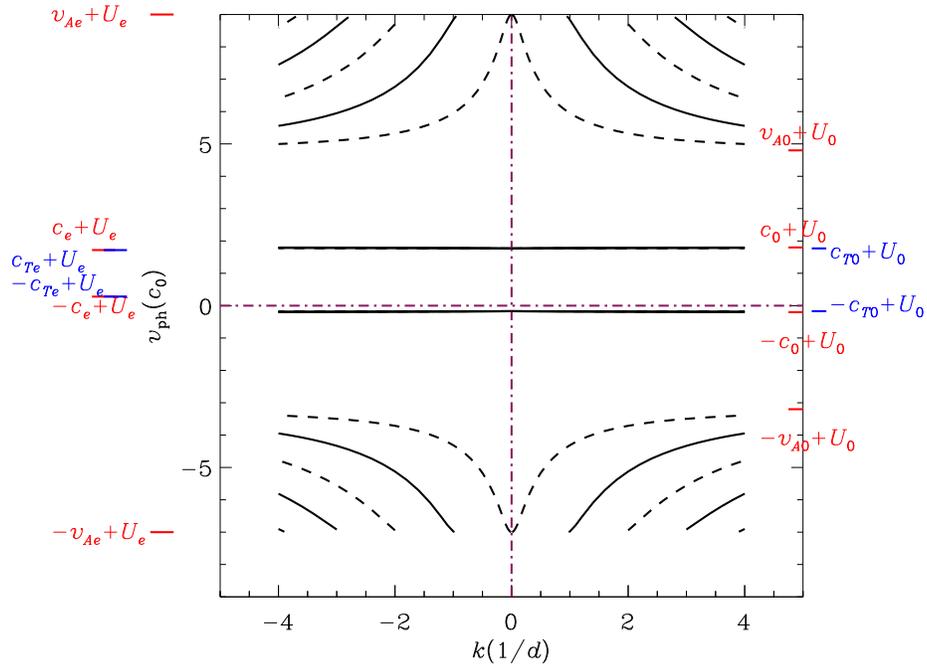}
\caption{Phase speeds $\vph$ as a function of longitudinal wavenumber $k$ in a general case where neither $U_0$ nor $U_e$ is zero.
The characteristic speeds are the same as adopted in Fig.\ref{fig_Cor_DR}, namely, 
   $v_{Ae}=8$, $c_e=0.72$, $v_{A0}=4$, and $c_0 = 1$. 
For $U_e$ and $U_0$, we choose them to be $c_0$ and $0.8 c_0$, respectively.
On the left (right) of the figure, the Doppler-shifted characteristic speeds external (interior) to the slab are given
   by the horizontal bars.
Kink and sausage modes are presented by the dashed and solid curves, respectively.
The horizontal and vertical dash-dotted lines in purple split the graph into four quadrants.
Note the symmetry between the left and right halves of the graph with respect to the vertical purple line, i.e., the $\vph-$ axis.
Actually this symmetry is valid for arbitrary $U_0$ and $U_e$.
}
 \label{fig_Cor_DR_NonZeroUe}
\end{figure}

\begin{figure}
\centering
\epsscale{.8}
\plotone{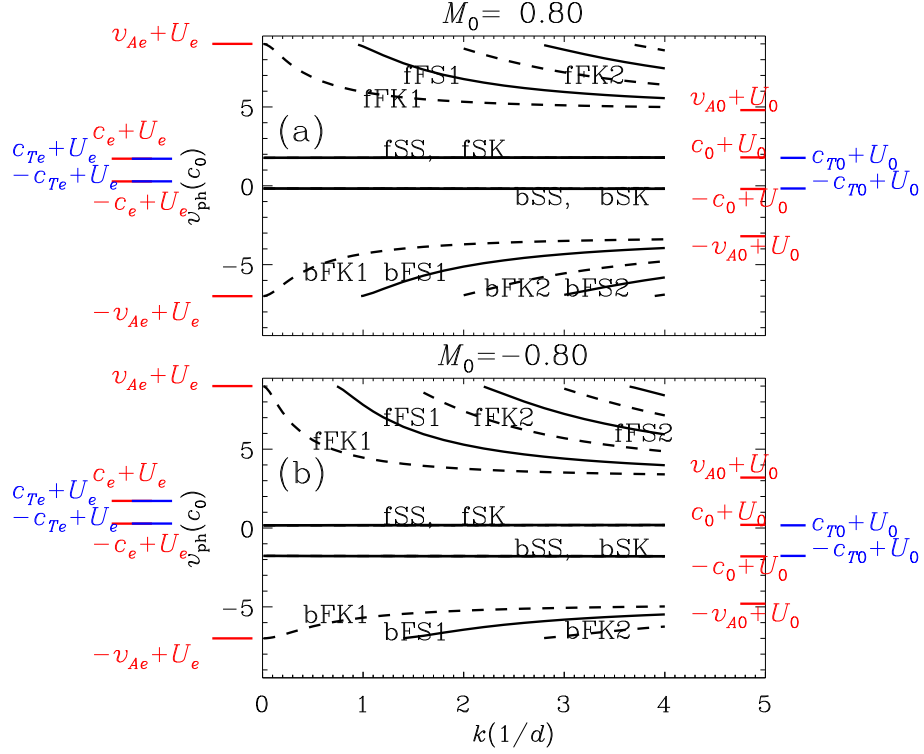}
\caption{Phase speeds $\vph$ as a function of longitudinal wavenumber $k$ where $U_e$ is non-zero.
The characteristic speeds are the same as adopted in Fig.\ref{fig_Cor_DR}, namely, 
   $v_{Ae}=8$, $c_e=0.72$, $v_{A0}=4$, and $c_0 = 1$. 
Panels (a) and (b) correspond to a $U_0$ of $0.8 c_0$ and $-0.8 c_0$, respectively.
In both panels, $U_e = c_0$.
On the left (right) of the figure, the Doppler-shifted characteristic speeds external (interior) to the slab are given
   by the horizontal bars.
Kink and sausage modes are presented by the dashed and solid curves, respectively.
The naming convention for the wave modes is the same as in Fig.\ref{fig_Cor_DR}.
}
 \label{fig_Cor_DR_NonZeroUe_PMU0}
\end{figure}

\begin{figure}
\centering
\epsscale{.8}
\plotone{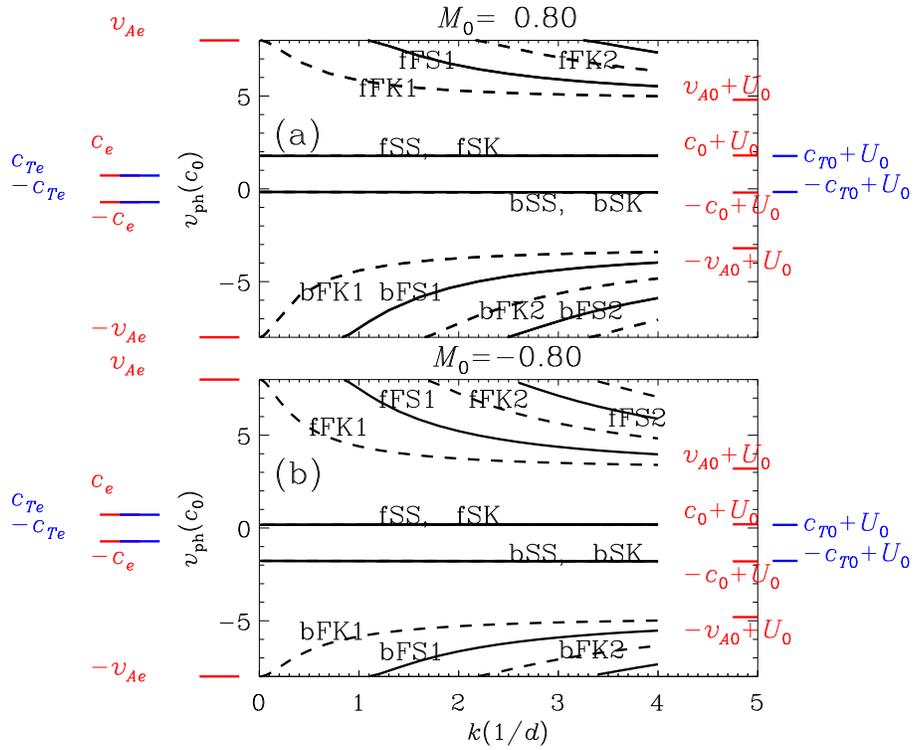}
\caption{Similar to Fig.\ref{fig_Cor_DR_NonZeroUe_PMU0} but for the case where $U_e=0$.
Note the symmetry that maps a curve in the 1st quadrant in (a), the one labeled fFK1 for instance,
    to a curve in the 4th quadrant in (b), the one labeled bFK1 in this example.
Likewise, this symmetry maps the curves in the 4th quadrant in (a) to the ones in the 1st quadrant in (b).
This symmetry takes place only when $U_e$ is zero. 
}
 \label{fig_Cor_DR_ZeroUe_PMU0}
\end{figure}

\begin{figure}
\centering
\epsscale{.9}
\plotone{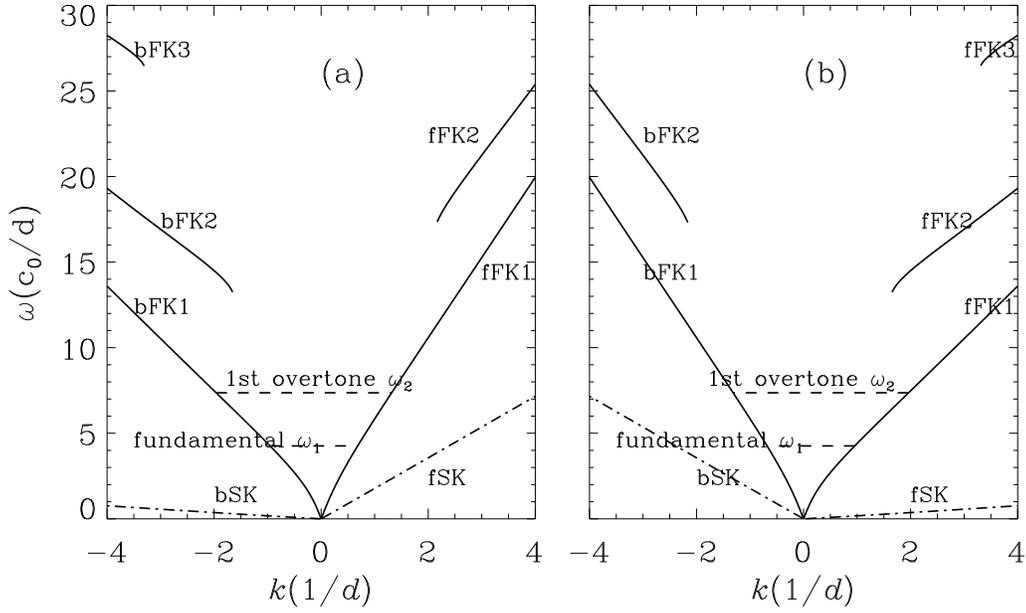}
 \caption{Angular frequency $\omega$ as a function of $k$ for propagating kink waves supported by a slab embedded in a
    static medium ($U_e = 0$).
 Panels (a) and (b) correspond to a $U_0$ of $0.8 c_0$ and $-0.8 c_0$, respectively.
Note the symmetry between (a) and (b), which is inherited from Fig.\ref{fig_Cor_DR_ZeroUe_PMU0}.
Consequently, this symmetry happens only when $U_e = 0$.
 }
 \label{fig_OmegaK_ZeroUe}
\end{figure}

{  
\begin{center}
{\bf APPENDIX}
\end{center}

\appendix
\section{ISSUES RELATED TO SYMMETRY PROPERTIES OF THE DISPERSION DIAGRAMS}

In this appendix, we examine the two symmetry properties presented in section~\ref{sec_DR_Oview} in some detail. In particular, we show that for the purpose of examining the period ratios of standing modes supported by a flowing structure embedded in a static background, one needs to consider only positive longitudinal wavenumbers $k$ and positive internal flow speed $U_0$.

Figure~\ref{fig_Cor_DR_NonZeroUe} illustrates what happens in a general case where neither $U_0$ nor $U_e$ is zero.
It presents the wave phase speed $\vph$ as a function of longitudinal wavenumber $k$, in a
   format similar to Figs.\ref{fig_Cor_DR} and~\ref{fig_Pho_DR}.
Note that both positive and negative $k$-s are considered here.
Kink and sausage waves are given by the dashed and solid curves. 
The horizontal and vertical dash-dotted lines in purple separate the graph into four quadrants.
The parameters used to construct this graph are identical to those adopted in Fig.\ref{fig_Cor_DR}b, except that now
   the external medium is not static, $U_e = c_0$ to be specific.
This rather arbitrary choice of $U_e$ is to illustrate symmetry property 1, which in fact holds for
   arbitrary values of $U_e$ and $U_0$.
Graphically, this property translates into the symmetry between the left half-plane (the 2nd and 3rd quadrants) 
   and the right one (the 1st and 4th quadrants).
In this sense one needs only to consider the 1st and 4th quadrants as long as one is interested only in trapped waves.

The second symmetry property is best illustrated if one compares Fig.\ref{fig_Cor_DR_NonZeroUe_PMU0}
   with Fig.\ref{fig_Cor_DR_ZeroUe_PMU0},
   where $\vph$ as a function of $k$ is shown for both a positive internal flow
   (panel a) and a negative one (panel b) with $|U_0| = 0.8 c_0$.
The characteristic speeds are identical to those adopted when constructing Fig.\ref{fig_Cor_DR}.   
The difference between Figs.\ref{fig_Cor_DR_NonZeroUe_PMU0}
   and \ref{fig_Cor_DR_ZeroUe_PMU0} is that in the former $U_e$ is not zero ($U_e$ chosen to be $c_0$ for illustrative purpose),
   whereas in the latter $U_e = 0$.
To label the waves, the same convention as in Figs.\ref{fig_Cor_DR} and~\ref{fig_Pho_DR} is adopted.
Note that in this coronal case no surface modes are introduced by the non-zero external flow, regardless of the sign of 
    the internal flow.
The body waves can, as in the case where $U_e=0$, be grouped into slow and fast ones.
The phase speed for the former is bounded by $\pm c_{T0}+U_0$ and $\pm c_0 + U_0$,
    and $\vph$ for the latter is bounded by $\pm v_{Ae}+U_e$ and $\pm v_{A0}+U_0$.    
The most important difference, as far as the symmetry properties are concerned, is that 
    Fig.\ref{fig_Cor_DR_NonZeroUe_PMU0} lacks a symmetry between the two panels, whereas in Fig.\ref{fig_Cor_DR_ZeroUe_PMU0}
    the curves in the 1st (4th) quadrant of panel (a)
    map to those in the 4th (1st) quadrant of panel (b).
For instance, the curve labeled bFK1 in Fig.\ref{fig_Cor_DR_ZeroUe_PMU0}a corresponds to fFK1 in Fig.\ref{fig_Cor_DR_ZeroUe_PMU0}b, 
    and likewise fFK2 in Fig.\ref{fig_Cor_DR_ZeroUe_PMU0}a corresponds to bFK2 in Fig.\ref{fig_Cor_DR_ZeroUe_PMU0}b.
As a matter of fact, this symmetry was present already in Figs.4 and 5 in \citet{2003SoPh..217..199T} where wave modes supported by a flowing
    cylinder is examined.
This is not surprising given that the authors also examined 
    the case where the external background is static,
    and that the $U_0$ they adopted in constructing Fig.4 ($U_0 = -0.25 v_{A0}$)
    is close to the one used for constructing Fig.5 ($U_0 = 0.35 v_{A0}$).
A perfect symmetry would result if the authors chose a pair of $U_0$ of the same magnitude but with opposite signs.     

The second symmetry, which exists only in the absence of external flows, does not invalidate the concept of 
    resonant flow instability (RFI)~\citep{Ryutova88,1990ApJ...349..335H,1996A&A...313..664E}
    (see \citeauthor{2011SSRv..158..289G}~\citeyear{2011SSRv..158..289G}
     and \citeauthor{2011SSRv..158..505T}~\citeyear{2011SSRv..158..505T} for recent reviews).
This point can be illustrated if one examines in detail, say, \citet{2000ApJ...531..561A} where the authors address the possible instability
    a velocity shear between a coronal plume and its environment may introduce. 
The authors choose to work in the zero-beta limit, and in a frame in which $U_e=0$ and $U_0$ is negative.
When a continuous Alfv\'en speed profile connects the internal and external values, RFI sets in when the ``originally forward-propagating''
    (in the sense that $\vph$ is positive when $U_0 = 0$) kink waves reach the continuum 
    of the originally backward-propagating Alfv\'en waves.
This happens at a $U_0$ smaller in magnitude than required by the non-resonant Kelvin-Helmholtz instability, 
    the difference between the two thresholds being quite significant 
    for high density contrasts between the internal and external media.
In light of the symmetry property we discussed, what Fig.3 in~\citet{2000ApJ...531..561A} shows for a negative $U_0$ can help one
    to foresee what happens when $U_0$ is positive.
In that case, RFI will also happen when $U_0$ reaches exactly the same magnitude, for at this point the originally backward-propagating
   kink waves reach the continuum of the originally forward-propagating Alfv\'en waves. 
Extending the present study to address RFI is physically very appealing, but may require quite some effort given the
   complexity related to solving eigenvalue problems that involve  
   dealing with the dissipative layers at the Alfv\'en resonances and the possible cusp resonances.

Figure~\ref{fig_OmegaK_ZeroUe} shows the wave angular frequency $\omega$ as a function of $k$, 
    constructed from the $\vph - k$ diagram presented in Fig.\ref{fig_Cor_DR_ZeroUe_PMU0}.
Here Figs.\ref{fig_OmegaK_ZeroUe}a and \ref{fig_OmegaK_ZeroUe}b correspond to $U_0 = 0.8 c_0$ and $-0.8 c_0$, respectively.
Only kink waves are presented, for otherwise the graph will be too crowded.
The first thing to note is that if one insists on $\omega$ being positive, one would need the curves in the 3rd quadrant of a $\vph-k$ diagram
    where both $k$ and $\vph$ are negative.
However, from symmetry property 1 illustrated in Fig.\ref{fig_Cor_DR_NonZeroUe}, one can construct the 3rd quadrant curves using
    the curves in the 4th quadrant
    by keeping $\vph$ unchanged while reversing the sign of $k$.
Second, Fig.\ref{fig_OmegaK_ZeroUe} inherits the same symmetry from Fig.\ref{fig_Cor_DR_ZeroUe_PMU0}.
For instance, the branch labeled bFK1 in Fig.\ref{fig_OmegaK_ZeroUe}a maps to the one labeled fFK1 in Fig.\ref{fig_OmegaK_ZeroUe}b.
Likewise, fFK1 in Fig.\ref{fig_OmegaK_ZeroUe}a corresponds to bFK1 in Fig.\ref{fig_OmegaK_ZeroUe}b.
Recall that to compute the period ratios for a given aspect ratio $d/L$, we simply choose the two curves corresponding to the propagating waves
   that are to combine to form a standing mode, and then measure $\omega_1$ and $\omega_2$ that correspond to the two horizontal cuts
   resulting in a separation being $2\pi d/L$ and twice that, respectively.
It follows from this procedure that the period ratio thus calculated would be the same in Figs.\ref{fig_OmegaK_ZeroUe}a
   and~\ref{fig_OmegaK_ZeroUe}b.
Evidently, if we choose to work in a frame where the external medium is not static, then the symmetry between
   Figs.\ref{fig_OmegaK_ZeroUe}a and~\ref{fig_OmegaK_ZeroUe}b
   will be broken,
   meaning that the period ratio for the same $d/L$ derived for a positive $U_0$ will differ from that derived for a negative one,
   even if the magnitudes of $U_0$ are equal.
This means, at any given non-zero $U_e$,  for one to examine the effects of $U_0$ on the period ratios
   one would have to first calculate a full range of different $U_0$, both positive and negative, 
   then construct the $\omega-k$ diagram, and finally measure $\omega_1$ and $\omega_2$.
This certainly merits a detailed discussion, but is beyond the scope of the present study.   
}

\clearpage

\clearpage

\clearpage

\clearpage

\clearpage

\clearpage

\clearpage

\clearpage

\clearpage

\clearpage

\clearpage
 
\clearpage
 
\clearpage


\begin{thebibliography}{}


\bibitem[Andries \& Goossens(2002)]{2002PhPl....9.2876A} 
  Andries, J., \& Goossens, M.\ 2002, Physics of Plasmas, 9, 2876 
\bibitem[Andries et al.(2000)]{2000ApJ...531..561A} 
  Andries, J., Tirry, W.~J., \& Goossens, M.\ 2000, \apj, 531, 561 

\bibitem[Andries et al.(2005)]{2005ApJ...624L..57A} Andries, J., Arregui,
I., \& Goossens, M.\ 2005, \apjl, 624, L57

\bibitem[Andries et al.(2009)]{2009SSRv..149....3A} Andries, J., van
Doorsselaere, T., Roberts, B., et al.\ 2009, \ssr, 149, 3

\bibitem[Antolin
\& Verwichte(2011)]{2011ApJ...736..121A} Antolin, P., \& Verwichte, E.\ 2011, \apj, 736, 121

\bibitem[Arregui et al.(2012)]{2012LRSP....9....2A} Arregui, I., Oliver,
R., \& Ballester, J.~L.\ 2012, Living Reviews in Solar Physics, 9, 2


\bibitem[Aschwanden(2004)]{2004psci.book.....A} Aschwanden, M.~J.\ 2004,
Physics of the Solar Corona, Springer Praxis

\bibitem[Aschwanden et al.(1999)]{1999ApJ...520..880A} Aschwanden, M.~J.,
Fletcher, L., Schrijver, C.~J., \& Alexander, D.\ 1999, \apj, 520, 880


\bibitem[Banerjee et al.(2007)]{2007SoPh..246....3B} Banerjee, D., 
Erd{\'e}lyi, R., Oliver, R., \& O'Shea, E.\ 2007, \solphys, 246, 3 


\bibitem[De Moortel(2006)]{2006RSPTA.364..461D} De Moortel, I.\ 2006, Royal
Society of London Philosophical Transactions Series A, 364, 461

\bibitem[De Moortel(2009)]{2009SSRv..149...65D} De Moortel, I.\ 2009, \ssr,
149, 65


\bibitem[De Moortel
\& Brady(2007)]{2007ApJ...664.1210D} De Moortel, I., \& Brady, C.~S.\ 2007, \apj, 664, 1210

\bibitem[De Moortel 
\& Nakariakov(2012)]{2012RSPTA.370.3193D} De Moortel, I., \& Nakariakov, V.~M.\ 2012, Royal Society of London Philosophical Transactions Series A, 370, 3193 

\bibitem[De Pontieu et al.(2007)]{2007Sci...318.1574D} De Pontieu, B.,
McIntosh, S.~W., Carlsson, M., et al.\ 2007, Science, 318, 1574

\bibitem[Dorotovi{\v c} et al.(2008)]{2008IAUS..247..351D} Dorotovi{\v c}, 
I., Erd{\'e}lyi, R., \& Karlovsk{\'y}, V.\ 2008, IAU Symposium, 247, 351 


\bibitem[Edwin \& Roberts(1982)]{1982SoPh...76..239E} Edwin, P.~M., \& Roberts, B.\ 1982, \solphys, 76, 239 

\bibitem[Edwin \& Roberts(1983)]{1983SoPh...88..179E} Edwin, P.~M., \& Roberts, B.\ 1983, \solphys, 88, 179 

\bibitem[Erd{\'e}lyi \& Fedun(2007)]{2007Sci...318.1572E} Erd{\'e}lyi, R., \& Fedun, V.\ 2007, Science, 318, 1572 

\bibitem[Erd{\'e}lyi \& Taroyan(2008)]{2008A&A...489L..49E} Erd{\'e}lyi, R., \& Taroyan, Y.\ 2008, \aap, 489, L49 

\bibitem[Erdelyi 
\& Goossens(1996)]{1996A&A...313..664E} Erdelyi, R., \& Goossens, M.\ 1996, \aap, 313, 664 

\bibitem[Erd{\'e}lyi \& Goossens(2011)]{2011SSRv..158..167E} Erd{\'e}lyi, R., \& Goossens, M.\ 2011, \ssr, 158, 167




\bibitem[Evans \& Roberts(1990)]{1990ApJ...348..346E} Evans, D.~J., \& Roberts, B.\ 1990, \apj, 348, 346 

\bibitem[Goossens et al.(2011)]{2011SSRv..158..289G} Goossens, M., 
Erd{\'e}lyi, R., \& Ruderman, M.~S.\ 2011, \ssr, 158, 289 

\bibitem[Harra et al.(2005)]{2005A&A...438.1099H} Harra, L.~K., D{\'e}moulin, P., Mandrini, C.~H., et al.\ 2005, \aap, 438, 1099

\bibitem[He et al.(2009)]{2009A&A...497..525H} He, J.-S., Tu, C.-Y., Marsch, E., et al.\ 2009, \aap, 497, 525


\bibitem[Hollweg et al.(1990)]{1990ApJ...349..335H} Hollweg, J.~V., Yang, 
G., Cadez, V.~M., \& Gakovic, B.\ 1990, \apj, 349, 335 



\bibitem[Inglis \& Nakariakov(2009)]{2009A&A...493..259I} Inglis, A.~R., \& Nakariakov, V.~M.\ 2009, \aap, 493, 259

\bibitem[Innes et al.(2003)]{2003SoPh..217..267I} Innes, D.~E., McKenzie, D.~E., \& Wang, T.\ 2003, \solphys, 217, 267

\bibitem[Jess et al.(2009)]{2009Sci...323.1582J} Jess, D.~B., Mathioudakis, M., Erd{\'e}lyi, R., et al.\ 2009, Science, 323, 1582

\bibitem[Joarder \& Narayanan(2000)]{2000A&A...359.1211J} Joarder, P.~S., \& Narayanan, A.~S.\ 2000, \aap, 359, 1211

\bibitem[Macnamara \& Roberts(2011)]{2011A&A...526A..75M} Macnamara, C.~K., \& Roberts, B.\ 2011, \aap, 526, A75




\bibitem[Marsh et al.(2009)]{2009ApJ...697.1674M} Marsh, M.~S., Walsh, R.~W., \& Plunkett, S.\ 2009, \apj, 697, 1674

\bibitem[Mathioudakis et al.(2012)]{2012SSRv..tmp...94M} Mathioudakis, M., 
Jess, D.~B., \& Erd{\'e}lyi, R.\ 2012, \ssr, 94 

\bibitem[Melnikov et al.(2005)]{2005A&A...439..727M} Melnikov, V.~F., Reznikova, V.~E., Shibasaki, K., \& Nakariakov, V.~M.\ 2005, \aap, 439, 727


\bibitem[Miteva et al.(2003)]{2003PhPl...10.4463M} Miteva, R., Zhelyazkov, I., \& Erd{\'e}lyi, R.\ 2003, Physics of Plasmas, 10, 4463

\bibitem[Morgan et al.(2004)]{2004ApJ...605..521M} Morgan, H., Habbal, S.~R., \& Li, X.\ 2004, \apj, 605, 521


\bibitem[Morton \& Erd{\'e}lyi(2009a)]{2009A&A...502..315M} Morton, R.~J., \& Erd{\'e}lyi, R.\ 2009a, \aap, 502, 315

\bibitem[Morton \& Erd{\'e}lyi(2009b)]{2009ApJ...707..750M} Morton, R.~J., \& Erd{\'e}lyi, R.\ 2009b, \apj, 707, 750

\bibitem[Morton \& Ruderman(2011)]{2011A&A...527A..53M} Morton, R.~J., \& Ruderman, M.~S.\ 2011, \aap, 527, A53

\bibitem[Morton et al.(2011)]{2011ApJ...729L..18M} Morton, R.~J., 
Erd{\'e}lyi, R., Jess, D.~B., \& Mathioudakis, M.\ 2011, \apjl, 729, L18 

\bibitem[Morton et al.(2012)]{2012NatCom..3.1315M}
Morton, R.J. et al.\ 2012, Nat. Commun., 3:1315, doi: 10.1038/ncomms2324





\bibitem[Nakariakov \& Roberts(1995)]{1995SoPh..159..213N} Nakariakov, V.~M., \& Roberts, B.\ 1995, \solphys, 159, 213

\bibitem[Nakariakov \& Ofman(2001)]{2001A&A...372L..53N} Nakariakov, V.~M., \& Ofman, L.\ 2001, \aap, 372, L53

\bibitem[Nakariakov \& Verwichte(2005)]{2005LRSP....2....3N} Nakariakov, V.~M., \& Verwichte, E.\ 2005, Living Reviews in Solar Physics, 2, 3

\bibitem[Nakariakov \& Erd{\'e}lyi(2009)]{2009SSRv..149....1N} Nakariakov, V.~M., \& Erd{\'e}lyi, R.\ 2009, \ssr, 149, 1


\bibitem[Nakariakov et al.(1999)]{1999Sci...285..862N} Nakariakov, V.~M., Ofman, L., Deluca, E.~E., Roberts, B., \& Davila, J.~M.\ 1999, Science, 285, 862
\bibitem[Nakariakov et al.(2003)]{2003A&A...412L...7N} Nakariakov, V.~M., Melnikov, V.~F., \& Reznikova, V.~E.\ 2003, \aap, 412, L7




\bibitem[Ofman \& Wang(2008)]{2008A&A...482L...9O} Ofman, L., \& Wang, T.~J.\ 2008, \aap, 482, L9


\bibitem[Roberts(2000)]{2000SoPh..193..139R} Roberts, B.\ 2000, \solphys, 193, 139

\bibitem[Roberts(2008)]{2008IAUS..247....3R} Roberts, B.\ 2008, IAU Symposium, 247, 3

\bibitem[Roberts et al.(1984)]{1984ApJ...279..857R} Roberts, B., Edwin, P.~M., \& Benz, A.~O.\ 1984, \apj, 279, 857

\bibitem[Ruderman(2010)]{2010SoPh..267..377R} Ruderman, M.~S.\ 2010, \solphys, 267, 377

\bibitem[Ruderman 
\& Roberts(2002)]{2002ApJ...577..475R} Ruderman, M.~S., \& Roberts, B.\ 2002, \apj, 577, 475 


\bibitem[Ruderman 
\& Erd{\'e}lyi(2009)]{2009SSRv..149..199R} Ruderman, M.~S., \& Erd{\'e}lyi, R.\ 2009, \ssr, 149, 199 

\bibitem[Ryutova(1988)]{Ryutova88}
Ryutova, M.~P. 1988, J. Exp. Theor. Phys., 94, 138



\bibitem[Srivastava et al.(2008)]{2008MNRAS.388.1899S} Srivastava, A.~K., Zaqarashvili, T.~V., Uddin, W., Dwivedi, B.~N., \& Kumar, P.\ 2008, \mnras, 388, 1899

\bibitem[Taroyan 
\& Ruderman(2011)]{2011SSRv..158..505T} Taroyan, Y., \& Ruderman, M.~S.\ 2011, \ssr, 158, 505 

\bibitem[Taroyan et 
al.(2007)]{2007A&A...462..331T} Taroyan, Y., Erd{\'e}lyi, R., Doyle, J.~G., \& Bradshaw, S.~J.\ 2007, \aap, 462, 331 


\bibitem[Terra-Homem et al.(2003)]{2003SoPh..217..199T} Terra-Homem, M., Erd{\'e}lyi, R., \& Ballai, I.\ 2003, \solphys, 217, 199


\bibitem[Terradas et al.(2011)]{2011ApJ...729L..22T} Terradas, J., Arregui, I., Verth, G., \& Goossens, M.\ 2011, \apjl, 729, L22


\bibitem[Tomczyk et al.(2007)]{2007Sci...317.1192T} Tomczyk, S., McIntosh, S.~W., Keil, S.~L., et al.\ 2007, Science, 317, 1192

\bibitem[Uchida(1970)]{1970PASJ...22..341U} Uchida, Y.\ 1970, \pasj, 22, 341 

\bibitem[Van Doorsselaere et al.(2007)]{2007A&A...473..959V} Van Doorsselaere, T., Nakariakov, V.~M., \& Verwichte, E.\ 2007, \aap, 473, 959

\bibitem[Van Doorsselaere et al.(2008a)]{2008ApJ...676L..73V} Van Doorsselaere, T., Nakariakov, V.~M., \& Verwichte, E.\ 2008a, \apjl, 676, L73

\bibitem[Van Doorsselaere et al.(2008b)]{2008A&A...487L..17V} Van Doorsselaere, T., Nakariakov, V.~M., Young, P.~R., \& Verwichte, E.\ 2008b, \aap, 487, L17

\bibitem[van Doorsselaere et al.(2009)]{2009A&A...508.1485V} Van Doorsselaere, T., Birtill, D.~C.~C., \& Evans, G.~R.\ 2009, \aap, 508, 1485

\bibitem[Van Doorsselaere et al.(2011a)]{2011ApJ...727L..32V} Van Doorsselaere, T., Wardle, N., Del Zanna, G., et al.\ 2011a, \apjl, 727, L32

\bibitem[Van Doorsselaere et al.(2011b)]{2011ApJ...740...90V} Van Doorsselaere, T., De Groof, A., Zender, J., Berghmans, D., \& Goossens, M.\ 2011b, \apj, 740, 90



\bibitem[Verwichte et al.(2004)]{2004SoPh..223...77V} Verwichte, E., Nakariakov, V.~M., Ofman, L., \& Deluca, E.~E.\ 2004, \solphys, 223, 77


\bibitem[Verwichte et al.(2009)]{2009ApJ...698..397V} Verwichte, E., Aschwanden, M.~J., Van Doorsselaere, T., Foullon, C., \& Nakariakov, V.~M.\ 2009, \apj, 698, 397

\bibitem[Verwichte et al.(2010)]{2010ApJ...717..458V} Verwichte, E., Foullon, C., \& Van Doorsselaere, T.\ 2010, \apj, 717, 458


\bibitem[Wang(2011)]{2011SSRv..158..397W} Wang, T.\ 2011, \ssr, 158, 397


\bibitem[White \& Verwichte(2012)]{2012A&A...537A..49W} White, R.~S., \& Verwichte, E.\ 2012, \aap, 537, A49

\bibitem[Zaitsev \& Stepanov(1975)]{ZS75}
Zaitsev, V.~V., \& Stepanov, A.~V.\ 1975, Issled. Geomagn., Aeronomii Fiz. Solntsa, 37, 3 
\end{thebibliography}
\end{document}